\documentclass{elsart}

\usepackage{graphicx}
\usepackage{amssymb}

\begin{document}

\begin{frontmatter}



\title{Dynamics of a nanoparticle as a one-spin system and beyond}

\author{H. Kachkachi}
\ead{kachkach@physique.uvsq.fr}

\address{Laboratoire de Magn\'{e}tisme et d'Optique,
Universit\'e de Versailles St. Quentin, \\
45 av. des Etats-Unis, 78035 Versailles, France}

\begin{abstract}
We review some recent results beyond the now established theory of
magnetization switching of a nanoparticle within the single-spin approximation.
The first extension is that of the Stoner-Wohlfarth model for magnetization static
switching under applied magnetic field including the effect of temperature
at long-time scales.
The second concerns a generalization of the N\'eel-Brown model for thermoactivated
dynamic magnetization switching to include the effect of exchange interaction in
the framework of Langer's theory in the intermediate-to-high damping limit.
We finally argue why the single-spin approximation is not appropriate for very
small nanoparticles.
\end{abstract}

\begin{keyword}
{\it nanomagnetism, magnetization dynamics, exchange interaction}
\PACS: 75.50.Tt - 75.10.Hk - 05.40.Jc
\end{keyword}
\end{frontmatter}

% main text
\section{\label{intro}Introduction}
Magnetic nanoparticles are very interesting and challenging systems to both
fundamental research in physics and technological applications.
From the point of view of fundamental physics, nanoparticles are interesting
because they offer a rich laboratory for the study of dynamic phenomena as
they show superparamagnetism at high temperature and exponentially slow
relaxation at low temperature.
On the other hand, due to their high coercivity, their magnetization enjoys
long-range stability, a very important property for the information-storage
technology.
In principle, particles of very small size may be used to increase the storage
density. However, this is hampered by many difficulties related with
the finite size of the particles.
In particular, surface and thermal effects become
dominant in such particles, and drastically affect the magnetization relaxation
time, and thereby the stability of the information stored in recording media.
Therefore, the aim of any study of such systems, be it experimental or
theoretical, is to understand the dynamic properties of small nanoparticles
taking account of spatial inhomogeneities due to, e.g., surface disorder.
For this purpose one has to tackle the difficult problem of including surface
effects in the calculation of the relaxation time of the particle
magnetization.
However, this requires a microscopic approach to account for
the local environment inside the particle, and thus include microscopic
interactions such as spin-spin exchange, together with magnetocrystalline
anisotropy and Zeeman interaction.
Unfortunately, this leads to a rather difficult task owing to the large
number of degrees of freedom which hinders any attempt to analyze the
energyscape.
For this reason, inter alia, calculations of the reversal time of the
magnetization of fine single-domain ferromagnetic particles initiated by
N\'{e}el \cite{nee49ancras}, and set firmly in the context of the theory of stochastic
processes by Brown \cite{bro63pr}, \cite{bro79ieee}, have invariably proceeded 
by ignoring all kind of interactions. Thus, the only terms which are taken into
account in these calculations are the internal magneto-crystalline anisotropy of
the particle, the random field due to thermal fluctuations, and the Zeeman term.
This assumes that the particle's atomic moments (or spins) switch in a
coherent manner, so that the magnetic state of the particle can be described by
a giant magnetic moment, this is the {\it one-spin approximation}, and the
particle is then dealt with as a one-spin system.

It is well known that the magnetization of a nanoparticle can overcome the
energy barrier, due to magneto-crystalline anisotropy, and thus reverse its
direction, at least in two ways \footnote{It has quite recently been shown
\cite{shuetal03prls}, experimentally and theoretically,
that efficient magnetization switching can be triggered by transverse field
pulses of a duration that is half the precession period.}: either under applied
magnetic field which suppresses the barrier, or by thermal effects which produce
statistical fluctuations.
Within the framework of the one-spin approximation,
the magnetization switching under applied magnetic field, at zero temperature,
is well described by the Stoner-Wohlfarth model \cite{stowoh48}.
This model has been confirmed by experiments on single cobalt particles by
Wernsdorfer et al. [see Ref.~\cite{wer01acp} for a review].
At finite temperature, but at long-time scales or quasi-equilibrium, this
switching occurs according to two regimes. At very low temperature, this is due
to the coherent rotation of all spins, as in the Stoner-Wohlfarth model, whereas
at higher temperature, the magnetization switches by changing its
magnitude~\footnote{ Rigorously, the switching of magnetization through a change
of its magnitude cannot, however, be explained within the single-spin
approximation [see section~\ref{sec:discussion}].}.
This results in a shrinking of the Stoner-Wohlfarth astroid as predicted by the
modified Landau theory \cite{kacgar01physa}, and confirmed by experiments
\cite{wer01acp}.
At short-time scales, and within the one-spin approximation, crossing of the
energy barrier activated by thermal energy is described by the N\'eel-Brown
model \cite{nee49ancras}, \cite{bro63pr}, \cite{bro79ieee} and its extensions [see
section~\ref{sec:NB}], again in agreement with experiments on individual
cobalt particles \cite{wer01acp}.

In both Stoner-Wohlfarth and N\'eel-Brown models, the energy of a
fine single-domain ferromagnetic nanoparticle only contains the internal
magneto-crystalline anisotropy and the Zeeman term.
The thermoactivated dynamics of the particle, e.g., within the stochastic
Langevin approach, are described by adding a stochastic field in the equations
of motion of the magnetization, in order to account for thermal effects exerted
on the particle by the surrounding bath. This stochastic field is
responsible for the statistical fluctuations of the magnetization direction.

In this paper we briefly describe these two models.
Then, we present two extensions thereof: the first is of the
Stoner-Wohlfarth model to include thermal effects on the magnetization switching
at long-time scales, or equivalently, at quasi-equilibrium.
The second extension is that of the N\'eel-Brown model which takes account of
exchange interaction and investigates its effect on the relaxation time of a
pair of magnetic moments, within Langer's approach.
It will be shown that the N\'eel-Brown result for the relaxation time of a rigid
magnetic moment is only recovered for very large exchange coupling, while for
weak coupling there appear new interesting features, giving a foretaste of the
intricacies and subtleties that should arise in a particle when treated as a
multi-spin system.

The paper is organized as follows:
In section~\ref{sec:osp} we deal with a nanoparticle in the single-spin
approximation. For such a single-spin system, we briefly discuss the
Stoner-Wohlfarth model describing the magnetization switching under the applied
magnetic field and then consider the effect of temperature within the framework
of Landau's theory. Next, the N\'eel-Brown model for
thermoactivated switching is reviewed.
We comment on the applicability of this model, and its further extensions to
non-axially symmetric potentials, to the calculation of the blocking temperature
of an assembly of nanoparticles.
In section~\ref{sec:hk_tsp}, we abandon the single-spin approximation and deal
with the simplest, though non trivial, problem of two exchange-coupled spins and
study the effect of the latter on the relaxation time within Langer's approach.
The results are then compared with those of the N\'eel-Brown model.
The last section motivates and discusses the generalization of the latter work
to a multi-spin particle.
\section{\label{sec:osp}Magnetization switching of a nanoparticle as a
single-spin system}
Before we proceed, we would like to stress the role of anisotropy.
Exchange energy is completely isotropic, that is it does not depend on the
direction in space in which the crystal is magnetized.
One often chooses the reference $z$ direction as the one along which the
component of the magnetization is calculated. However, this direction does not
really have any meaning in the limit of zero applied field.
In fact, due to global rotation the net magnetization vanishes in zero applied
field.
This not only contradicts experiments but it is also in conflict with everyday
experience that, for instance, the particles in an audio or video tape remain
magnetized, and do not lose the recorded information upon switching off the
writing field.
This is so because real magnetic materials are not isotropic, and that the
reference $z$ direction in question is actually defined by anisotropy.
There are several types of anisotropy, but the most common is the
magneto-crystalline anisotropy, caused by the spin-orbit interaction. Indeed,
the electron orbits are linked to the crystallographic structure, and by their
interaction with the spins they make the latter prefer to align along
well-defined crystallographic axes.
There are therefore some directions in space in which it is energytically more
favorable to magnetize a given crystal than in other directions, the difference
being given by the direction-dependent anisotropy energy.
As for the exchange interaction, quantitative estimates of the
spin-orbit interaction from basic principles are also possible but the accuracy
is not as good. Therefore, the anisotropy term is always written as a
phenomenological expression, which is actually the first term of a power series
expansion that take into account the crystal symmetry, and the corresponding
coefficient is usually taken from experiment.
The magneto-crystalline energy is usually $2-4$ orders of magnitude smaller than
exchange energy, but the direction of the magnetization is determined only by
anisotropy.
On the other hand, exchange coupling tends to align all the spins inside the
particle parallel to each other, while anisotropy tends to align them along a
certain crystallographic direction.
A compromise between exchange and anisotropy is obtained by aligning all spins
parallel to each other and to the anisotropy direction.

Other features of a magnetic material, namely its subdivision into
domains, is then caused but yet another energy term, called the magnetostatic
self-energy (demagnetization or shape energy).
This is the surface term that stems from the dipolar interactions. It is
irrelevant for spherical samples.
However, for many materials, slight deviation from spherical shape renders the
shape anisotropy very important especially for the spatial dependence of
the magnetization.
On the other hand, as shown by Brown and Morrish in \cite{bromor57pr}, a
single-domain particle with an arbitrary shape is equivalent to a suitably
chosen general ellipsoid of the same volume, as far as the total energy is
concerned.
In all the sequel, we shall assume that the particle under study is spherical
and single-domain.

\subsection{\label{sec:SW}Magnetization switching under
magnetic field: Stoner-Wohlfarth model}
\begin{figure}[floatfix]
\includegraphics[angle=-90, width=10cm]{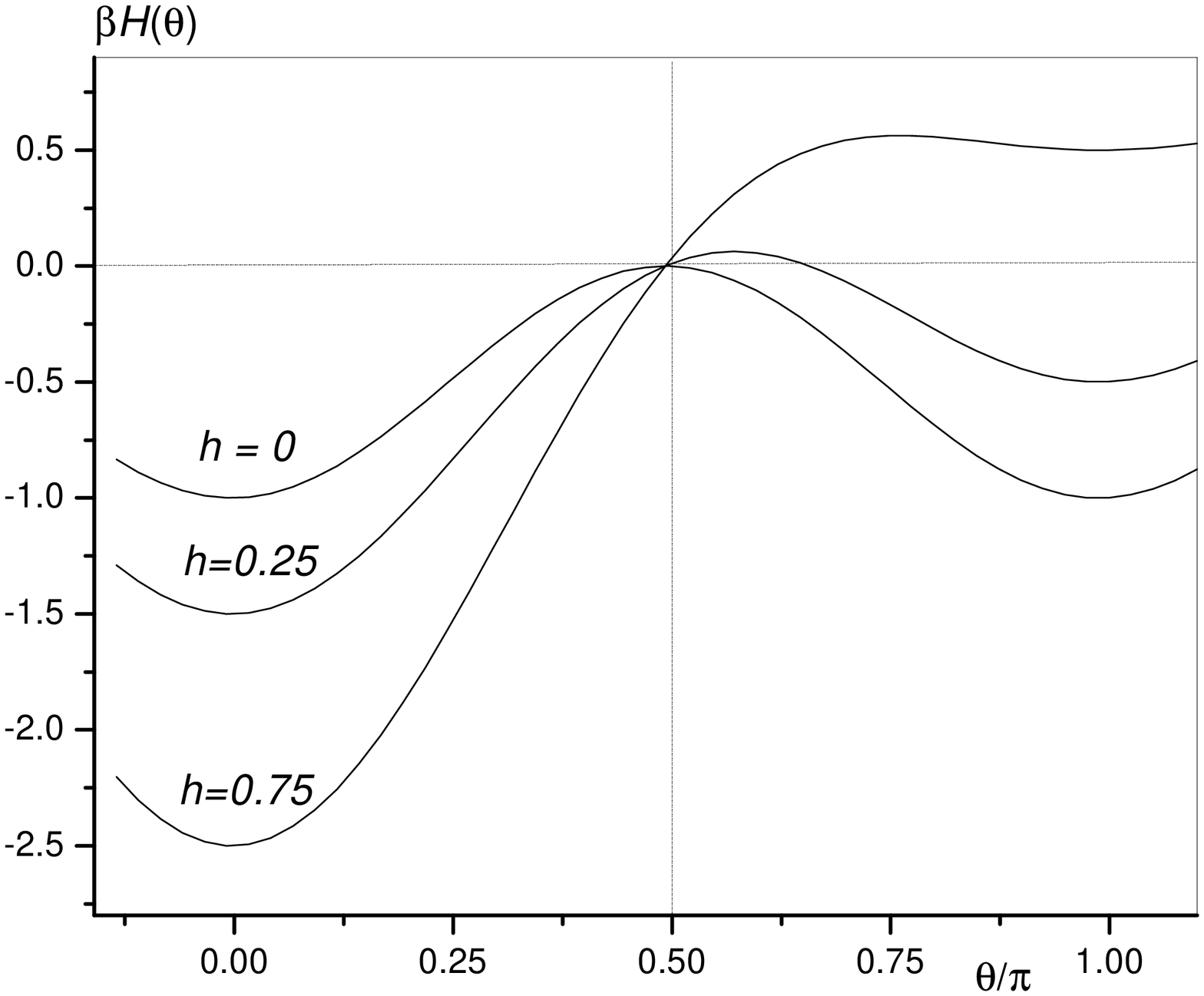}
\caption{\label{sw_landscape}Magnetic energy in the case of longitudinal field
for some values of the reduced field $h$. Upon increasing the field
the number of potential wells changes from two to one.}
\end{figure}

In small magnetic particles exchange interaction may be strong enough to hold
all spins tightly parallel to each other, and prevents spatial dependence of the
particle's magnetization.
We have shown \cite{kacgar01physa}, \cite{kacdim02prb} that this no longer
holds when free-boundary effects and/or surface anisotropy are taken into
consideration. %
Ignoring surface effects, or in other words spatial
inhomogeneities, the exchange energy is a constant and plays no role in the
energy minimization.
In the latter only the anisotropy energy and the interaction with the applied
field are relevant.
Accordingly, let us consider the simplest case of a magnetic moment ${\bf m}$
with a uniaxial anisotropy axis ${\bf e}$ along the $z$ direction of the applied
field.
The corresponding Hamiltonian reads
\begin{equation}\label{osp1}
{\mathcal H} = - \frac{Kv}{m^2}({\bf m}.{\bf e})^2 -{\bf m}.{\bf H},
\end{equation}
where $K$ is the magneto-crystalline anisotropy constant and $v$ is the volume
of the particle.
Upon writing ${\bf m}=m{\bf s}, {\bf H}=H{\bf e}_h$, and introducing the
dimensionless anisotropy and field parameters
\begin{equation}\label{anisfieldparams}
\sigma=\frac{Kv}{k_BT},\quad h=\frac{Hm}{2Kv},
\end{equation}
where $k_B$ is the Boltzmann constant, the Hamiltonian (\ref{osp1}), divided by
thermal energy, becomes
\begin{equation}\label{osp2}
-\beta{\mathcal H}=\sigma\left[({\bf s}.{\bf e})^2+2h({\bf s}.{\bf e}_h)\right].
\end{equation}
Denoting by $\theta$ the angle between the magnetization direction ${\bf s}$ and
the field direction ${\bf e}_h$, we write (\ref{osp2}) as
\begin{equation}\label{osp3}
-\beta{\mathcal H}=\sigma\left[\cos^2\theta+2h\cos\theta\right].
\end{equation}
For simplicity, we only consider the case of easy axis, i.e., $\sigma<0$. For
$\sigma>0$ (easy plane) the results are the same only that the energy maximum
and minimum are interchanged.
The extrema of ${\mathcal H}$ and their nature are obtained by setting to zero its
first derivative with respect to $\theta$, and evaluating the
second derivative at these extrema.
The results are presented in Table \ref{tab:extrema_osp}.
\begin{table}
\caption{\label{tab:extrema_osp}}
%\begin{ruledtabular}
\begin{tabular}{lll}
Field &\quad Minima &\quad Maxima\\
\hline
$|h|<1$ &\quad $\theta=0,\pi$ &\quad $\theta=\arccos(-h)$\\
$h>1$   &\quad $\theta=0$     &\quad $\theta=\pi$\\
$h<-1$  &\quad $\theta=\pi$   &\quad $\theta=0$\\
\end{tabular}
%\end{ruledtabular}
\end{table}
Thus, for $|h|<1$ the energy has minima at $\theta=0$ and $\theta=\pi$,
separated by a maximum at $\theta_m=\arccos(-h)$. For $|h|\gtrsim 1$ the upper
(also the shallower) energy minimum ($\theta=\pi$ for $h>0$) turns into a
maximum as it merges with the intermediate maximum at $\theta_m$,
which disappears (see Fig.~\ref{sw_landscape}).
%
%=========================================================================
%=========================================================================
%
%=========================================================================
\begin{figure}[h!]
\includegraphics[angle=-90, width=13cm]{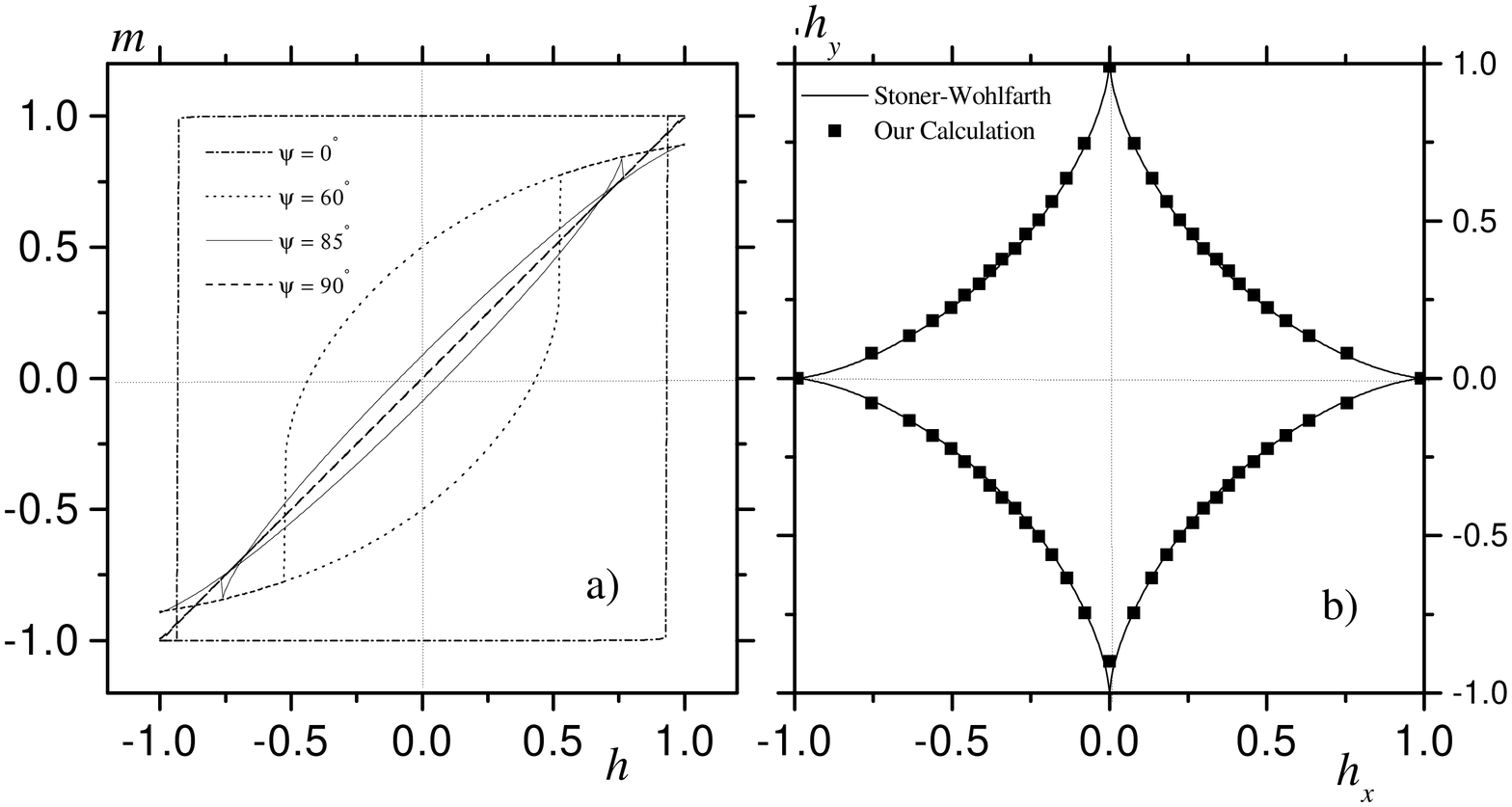}
\vspace{-2cm}
\caption{\label{sw_hystast}
Left: (numerical) hysteresis loops for different values of $\psi $ increasing
inwards: $\psi =0,60^{\circ},85^{\circ},90^{\circ}$, for a $3^{3}$
particle with uniaxial anisotropy and strong exchange coupling. Right:
(numerical in squares and analytical in full line) SW astroid for
the same particle.}
\end{figure}
%=========================================================================
%
From the values of the energy at $\theta=0,\pi$ and, when it exists, at the
intermediate maximum $\theta_m$, one obtains the energy-barrier heights
($|h|<1$)
\begin{equation}\label{barrier_height}
\Delta {\mathcal H} = \beta\left[{\mathcal H}(\theta_m)-{\mathcal H}(0,\pi)\right] =
\sigma(1\pm h)^2.
\end{equation}

The analysis above helps solve for the hysteresis curve, an
important part of magnetism, for small ferromagnetic particles. Such
calculations are known as the Stoner-Wohlfarth model \cite{stowoh48}.
In their original study, Stoner and Wohlfarth considered shape anisotropy
instead of uniaxial anisotropy. However, their model is usually applied to the
latter.
In addition, they also considered the more general situation of a field applied
at an arbitrary angle $\psi$ to the easy axis.
From the above analysis we see that there is a unique minimum for $|h|>1$
while there are two minima for $|h|<1$. This is due to the multivaluedness of
the trigonometric functions entering the Hamiltonian (\ref{osp3}) and its
derivatives with respect to $\theta$.
In order to obtain a unique solution, one has to specify and follow the history
of the field $h$ for each angle $\psi$. One usually starts at saturation and
increases (or decreases) the field across zero until it reaches the value at
which the energy barrier disappears. This field marks the {\it limit of
metastability} and is called the {\it critical field}. In Stoner-Wohlfarth
model, this is given by $|h|=1$ [for $\psi=0$].
For arbitrary $\psi$ the critical field is given by
$$
h_c = \frac{1}{\left(\cos^{2/3}\psi + \sin^{2/3}\psi\right)^{3/2}}.
$$
One can also define what is called the {\it switching field}, that is the field
at which the magnetization changes sign.

In Fig.~\ref{sw_hystast} we present (on the left) the hysteresis loop at
different angles $\psi$, and on the right (in full line) the angular dependence
of the switching field, the so-called Stoner-Wohlfarth astroid, as obtained from
the Stoner-Wohlfarth model [see Eq.~(\ref{swasrtsmall}) with $a=0$].
In fact, the hysteresis loops in Fig.~\ref{sw_hystast} and the astroid
in squares, were obtained from the numerical solution of the Landau-Lifshitz
equation \cite{kacdim02prb}.

Stoner-Wohlfarth astroids for cubic anisotropy have been obtained by Thiaville
\cite{thi98jmmm00prb} using a geometrical approach.

\subsection{\label{sec:TESW}Thermal effects on the Stoner-Wohlfarth
model: Landau theory}
Before dealing with the effect of thermoactivation on the magnetization
switching between the different energy minima, and discuss the
calculation of the corresponding relaxation time, we now consider the
effect of temperature on the Stoner-Wohlfarth model \cite{kacgar01physa}.
More precisely, we consider magnetization switching at a long-time scale, i.e.,
at quasi-equilibrium.
In \cite{kacgar01physa} we derived a free energy for weakly
anisotropic ferromagnets which is valid in the whole range of temperature and
interpolates between the micromagnetic energy at zero temperature and the Landau
free energy near the Curie point $T_c$.
This free energy takes into account the change of the magnetization magnitude
due to thermal effects, in particular, in the inhomogeneous states.
As an illustration, we studied the thermal effect on the Stoner-Wohlfarth
astroid and hysteresis loop of a ferromagnetic nanoparticle assuming that it is
in a single-domain state.
Within this model, the saddle point of the particle's free energy, as well as
the metastability boundary, are due to the change in the magnetization magnitude 
sufficiently close to $T_c$, as opposed to the usual homogeneous rotation
process at lower temperatures.

For weakly anisotropic magnets, where the anisotropy energy is much weaker than
the homogeneous exchange energy, the magnetization magnitude $M$ is either small
or only slightly deviates from $M_e$, the equilibrium magnetization at zero
field.
Thus the condition for the magnetization minimizing the free energy
derived in \cite{kacgar01physa}, $|M-M_e| \ll M_s$, where $M_s$ is
the saturation magnetization at $T=0$, is satisfied in the whole
temperature range.
In \cite{kacgar01physa} we presented the derivation of this free
energy and illustrated some of its features in the case of a single-domain
magnetic particle with a uniaxial anisotropy.

For single-domain magnetic particles in a homogeneous state, the gradient
(exchange) terms in the free energy can be dropped and the free energy can be
written in the form
\begin{eqnarray}\label{freduced}
&&
F = F_e + (VM_e^2/\chi_\perp) f \nonumber \\
&&
f=-{{\bf n} \cdot {\bf
h}}+\frac{1}{2}(n_{x}^{2}+n_{y}^{2})+\frac{1}{4a}(n^{2}-1)^{2},
\end{eqnarray}
where $V$ is the particle's volume and $f$ the reduced free energy in
terms of the reduced variables
\begin{equation}\label{defnha}
{\bf n} \equiv {\bf M}/M_e,
\qquad {\bf h} \equiv {\bf H}\chi_\perp/M_e,
\qquad a \equiv 2\chi_\|/\chi_\perp.
\end{equation}
where $\chi_\|/$ and $\chi_\perp$ are the longitudinal and transverse
susceptibilities.
We see that the parameter $a$ here controls the rigidity of the magnetization
vector; it goes to zero in the zero-temperature limit (with fixed magnetization
modulus) and diverges at $T_c$.

For fields ${\bf h}$ inside the Stoner-Wohlfarth astroid [see
Fig.~\ref{sw_hystast} (right)], which will be generalized here to nonzero
temperatures, $f$ has two minima separated by a barrier.
Owing to the axial symmetry, one can set $n_y=0$ for the investigation of the
free energy scape.
The minima, saddle points, and the maximum can be found from the equations
$\partial f/\partial n_{x}=\partial f/\partial n_{z}=0$, or, explicitly
\begin{eqnarray}\label{eqextrema}
&&
n_{z}(n^2-1) = ah_z \nonumber \\
&&
n_{x}(n^2-1+a)=ah_x.
\end{eqnarray}
Solving for $n_x$ one obtains a 5th-order equation for
$n_z$
\begin{equation}\label{nzeq}
h_x^2 n_z^3 = (h_z+n_z)^2(ah_z + n_z-n_z^3).
\end{equation}

In zero field, the characteristic points of the energyscape can be simply
found from Eqs.~(\ref{eqextrema}).
One of these points is $n_x=n_z=0$, which is a local maximum for $a<1$ and a
saddle point for $a>1$.
The minima are given by $n_x=0$, $n_z=\pm 1$.
The saddle points correspond to $n_z=0$, while from the second of
Eqs.~(\ref{eqextrema}) one finds
\begin{eqnarray}\label{nxsaddle}
n_x =
\left\{
     \begin{array}{ll}
            \pm\sqrt{1-a},  &  a \leq 1 \\
            0,              &  a \geq 1.
     \end{array}
\right.
\end{eqnarray}
In fact, due to the axial symmetry, for $a<1$ one has  a saddle circle
$n_x^2+n_y^2=1-a$, rather than two saddle points.
The free-energy barrier following from this solution is given by
\begin{equation}\label{febarrier}
\Delta f \equiv f_{\rm sad} - f_{\rm min} =
\left\{
\begin{array}{ll}
(2-a)/4,          &   a \leq 1 \\
1/(4a),           &  a \geq 1.
\end{array}
\right.
\end{equation}
%
%=========================================================================
\begin{figure}[floatfix]
\includegraphics[angle=-90, width=7.5cm]{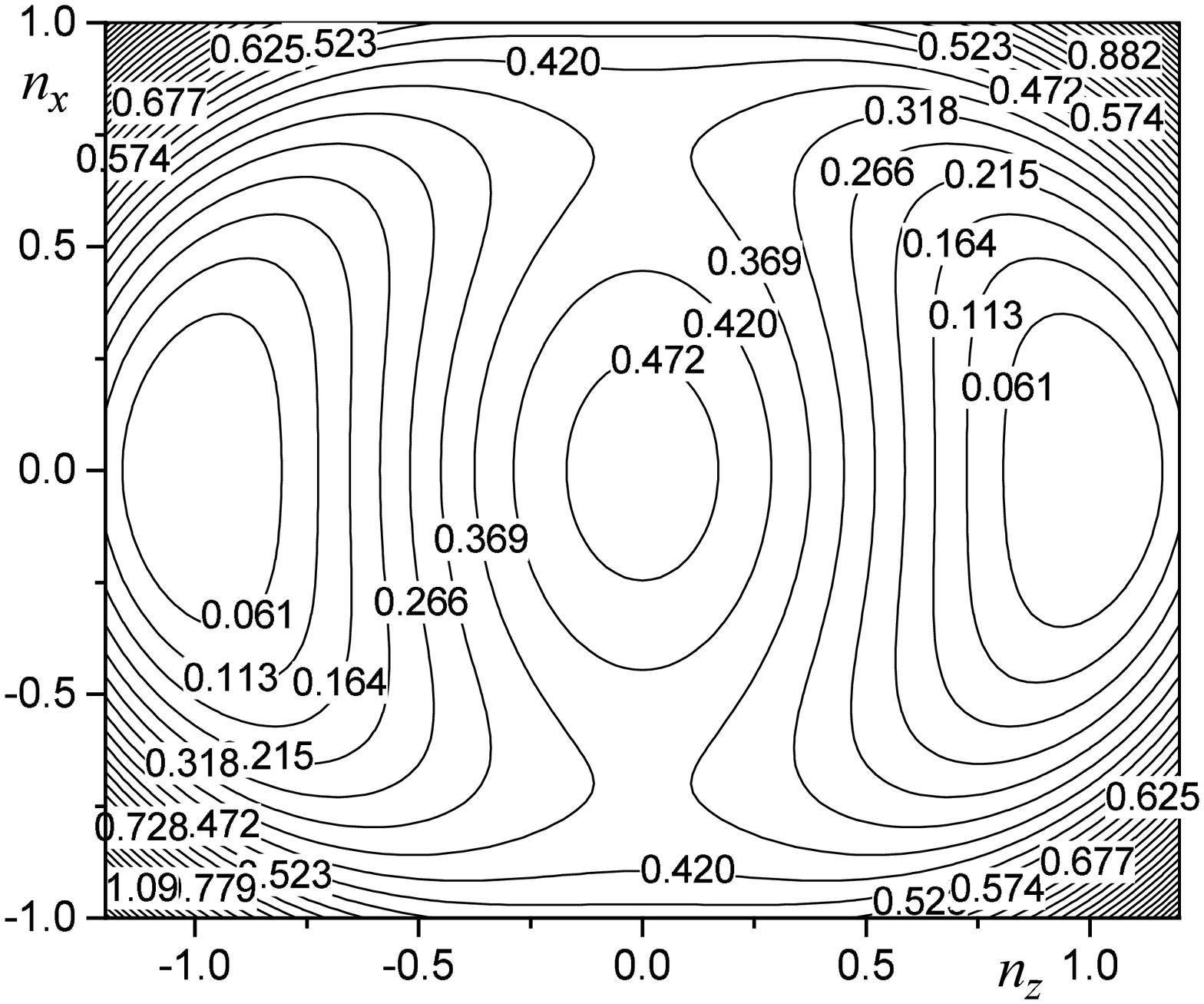}
\includegraphics[angle=-90, width=7.5cm]{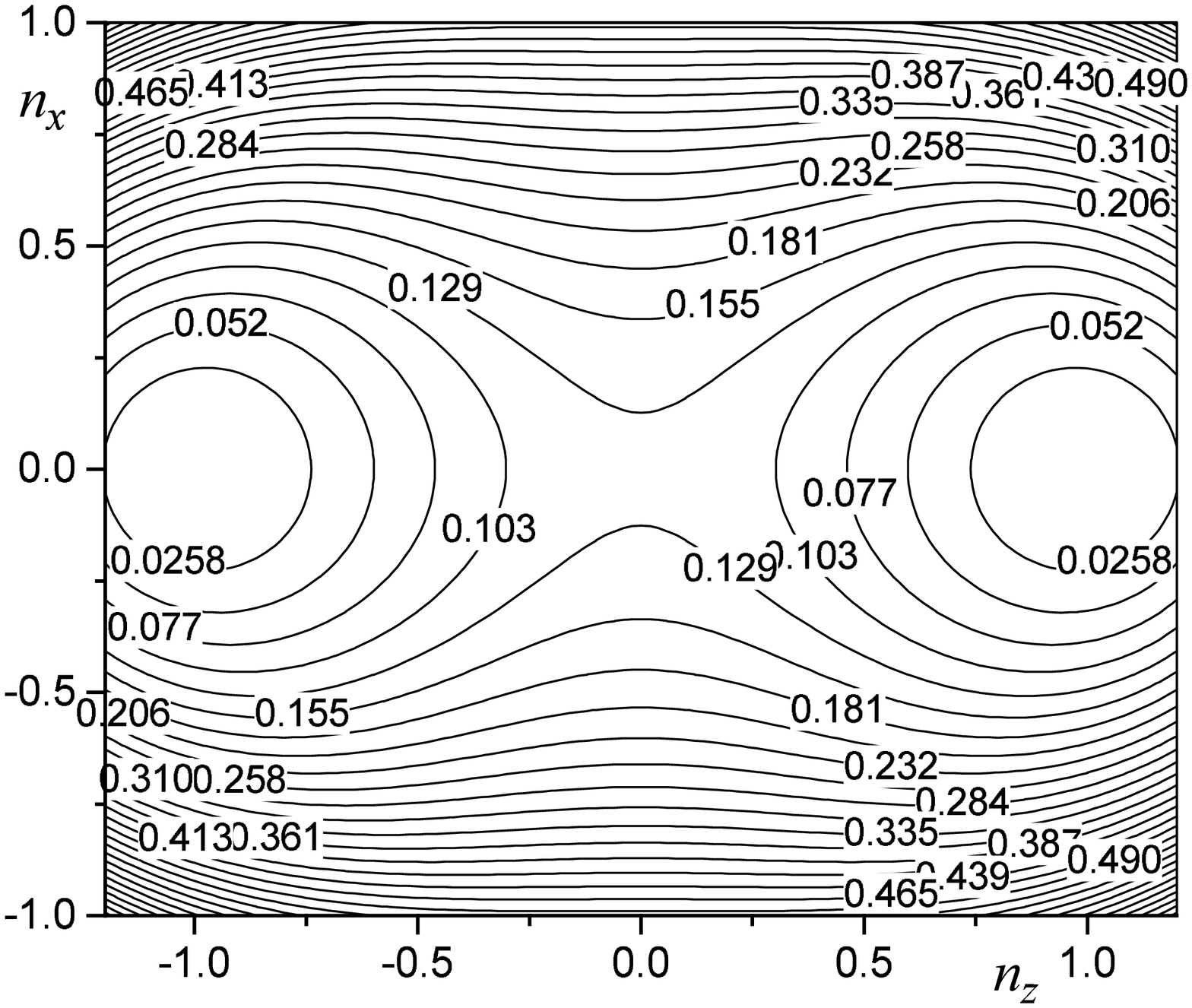}
\caption{\label{tsw_land}
The free energy of a ferromagnetic particle with uniaxial anisotropy
[$f$ in Eq.\ (\protect\ref{freduced}) in zero field] for $a\equiv
2\chi_\|/\chi_\perp = 0.5$ (left) and $a=2$ (right) corresponding to
lower and higher temperatures, respectively.}
\end{figure}
%=========================================================================
%

The free-energy landscape in zero field is shown in Fig.~\ref{tsw_land}.
At nonzero temperatures $a>0$, the magnitude of the magnetization at the saddle
is smaller than unity since it is directed perpendicular to the easy axis, and
for this orientation the ``equilibrium" magnetization is smaller than in the
direction along the $z$ axis.
For $a>1$, the two saddle points, or rather the saddle circle, degenerate into
a single saddle point at $n_x=n_z=0$, and the local maximum there disappears.
That is, for the magnetization to overcome the barrier, it is easier to change
its magnitude than its direction.
This is a phenomenon of the same kind as the phase transition in ferromagnets
between the Ising-like domain walls in the vicinity of $T_c$ (the magnetization
changes its magnitude and is everywhere directed along the $z$ axis) and the
Bloch walls at lower temperatures \cite{bulgin63jetp64jetp},
\cite{koegarharjah93}.

The Stoner-Wohlfarth curve separates the regions where there are two minima and
one minimum of the free energy.
On this curve the metastable minimum merges with the saddle point and loses its
local stability. The corresponding condition is
\begin{equation}\label{swcond}
\partial ^{2}f/\partial n_{x}^{2}\times \partial ^{2}f/\partial n_{z}^{2}
-(\partial ^{2}f/\partial n_{x}\partial n_{z})^{2} = 0,
\end{equation}
which upon using Eq.~(\ref{nzeq}) leads to the quartic equation for $n_z$
\begin{equation}\label{swcondnz}
h_z[(2+a)n_z + 3ah_z] + 2n_z^4=0.
\end{equation}
Before considering the general case, let us analyze the limiting cases $a\ll 1$
and $a\gg 1$.

At low temperatures, i.e., $a\ll 1$, the magnetization only slightly deviates
from its equilibrium value, and to first order in $a$, one obtains
\begin{equation}\label{nrel}
n^2 \cong 1 - an_z^2 \qquad {\rm or} \qquad
n_x^2 + (1+a) n_z^2 \cong 1.
\end{equation}
From Eq.~(\ref{swcondnz}) and the analogous equation for $n_x$ one derives
the equation for the Stoner-Wohlfarth astroid
\begin{equation}\label{swasrtsmall}
h_x^{2/3} + [(1+a/2)h_z]^{2/3}  \cong 1, \qquad a\ll 1.
\end{equation}
One can see that, in comparison with the standard zero-temperature
Stoner-Wohlfarth astroid, i.e. at $a=0$, $h_z$ is rescaled.
The critical field in the $z$ direction decreases because of the field
dependence of the magnetization magnitude at nonzero temperatures.

Similarly, in the case $a\gg 1$, i.e. near $T_c$, Eq.\ (\ref{swcond}) leads to
the equation for Stoner-Wohlfarth curve
\begin{equation}\label{nrel1}
n_x^2 + 3 n_z^2 \cong 1.
\end{equation}
and upon using Eq.~(\ref{eqextrema}) we infer
\begin{equation}\label{nznxonsw1}
n_z \cong -(ah_z/2)^{1/3}, \qquad n_x \cong h_x.
\end{equation}
Making use of this result in Eq.\ (\ref{nrel1}), one obtains another limiting
case of the Stoner-Wohlfarth curve
\begin{equation}\label{swalarge}
h_x^2 + 3(ah_z/2)^{2/3} \cong 1, \qquad a\gg 1.
\end{equation}
In this case, the critical field (i.e., the field on the Stoner-Wohlfarth curve)
in the $z$ direction is strongly reduced, and there is no singularity in the
dependence $h_{cz}(h_x)$ at $h_x=0$.
%
%=========================================================================
\begin{figure}[floatfix]
\includegraphics[angle=-90, width=14cm]{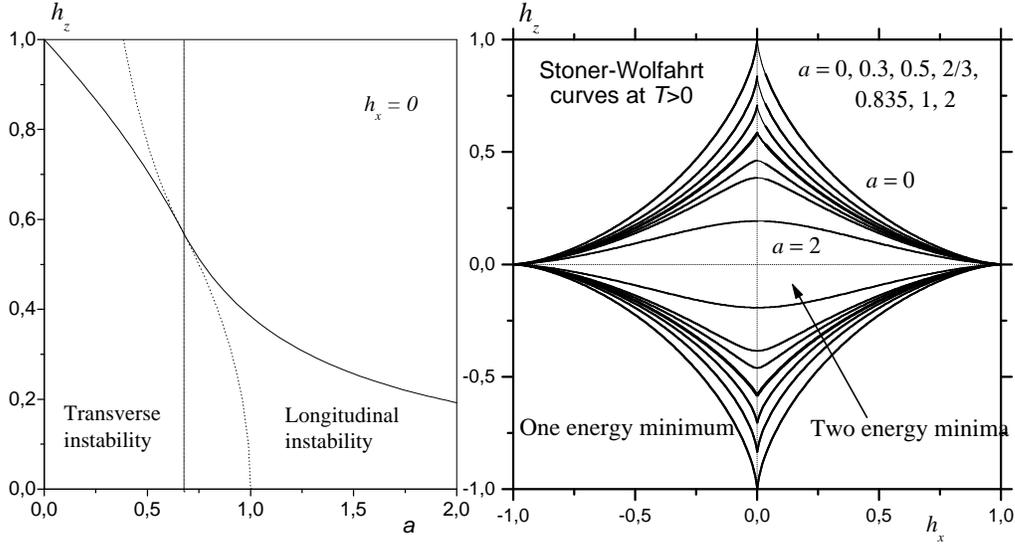}
\vspace{-1.5cm}
\caption{\label{tsw_hz}
Left: Dependence $h_z(a)$ at $h_x=0$ on the Stoner-Wohlfarth curve.
Right: The Stoner-Wohlfarth curves at different temperatures, $a\equiv
2\chi_\|/\chi_\perp=0$ ($T=0$), 0.3, 0.5, 2/3, 0.835, 1, 2.
}
\end{figure}
%=========================================================================
%

The qualitatively different character of the Stoner-Wohlfarth curves in these
two cases is due to the different mechanisms pertaining to the loss of the local
stability for the field applied along the $z$ axis.
For $h_x=0$ the mixed derivative $(\partial ^{2}f/\partial n_{x}\partial n_{z})$
vanishes and Eq.\ (\ref{swcond}) factorizes,
\begin{equation}\label{swcondfact}
(a-1+n_z^2)(-1+3n_z^2)=0.
\end{equation}
Vanishing of the first factor in this equation corresponds to the loss of
stability with respect to the rotation of the magnetization,
$\partial ^{2}f/\partial n_{x}^{2}=0$, and vanishing of the second factor,
$\partial ^{2}f/\partial n_{z}^{2}=0$, implies the loss of stability with
respect to the change of the magnetization magnitude.
Using $n_x=0$, with the help of the first equality in Eq.~(\ref{nzeq})
one obtains in the two cases
\begin{equation}\label{hzc}
h_{cz} =
\left\{
\begin{array}{ll}
h_{z\|} \equiv \sqrt{1-a},           & a \leq 2/3 \\
h_{z\perp} \equiv 2/(3^{3/2}a),      & a \geq 2/3.
\end{array}
\right.
\end{equation}
Note that the transition between the two regimes occurs here at a different
value of $a$ than in Eq.~(\ref{nxsaddle}).
The dependence $h_z(a)$ at $h_x=0$ is shown in Fig.~\ref{tsw_hz} (left).

In the general case, it is easier to find the Stoner-Wohlfarth curve
numerically from Eqs.~(\ref{eqextrema}) and (\ref{swcondnz}).
The results in the whole range of $a$ are shown in Fig.~\ref{tsw_hz} (right).

We have shown that, for single-domain magnetic particles with
uniaxial anisotropy, thermal effects qualitatively change
the free-energy landscape at sufficiently high temperatures, so that the passage
from one free-energy minimum to the other is realized by the {\em uniform change
of the magnetization magnitude} rather than the {\em uniform rotation}.
This also qualitatively changes the character of the Stoner-Wohlfarth curve and
hysteresis loops \cite{kacgar01physa}.
These effects cannot be observed with standard methods, however, because
keeping the height of the free-energy barrier much larger than thermal energy
requires so large particle sizes that the single-domain criterion is no longer
satisfied [see discussion in Sect.~6 of Ref.\cite{kacgar01physa}].
For the uniform states, the theory is valid at low temperatures, but then the
thermal effects considered here are small corrections to the
zero-temperature results.

Finally, we would like to mention that quite recently, experimental results have
been obtained by Jamet et al. \cite{jametal01prl} on 3 nm cobalt
nanoparticles which clearly show the disappearance of the singularity near $H_x
= 0$ at a temperature circa $8$ K (the blocking temperature being $14$ K).
The height of the experimental astroid decreases nearly as its width with
increasing temperature, but it does not become flat as predicted by our
calculations. This failure of our theory, in addition to the fact that this
effect is predicted at much higher temperatures, is not
surprising considering the fact that $T\ll T_c$.
Nevertheless, the disappearance of the singularity is definitely predicted by
our theory.

\subsection{\label{sec:NB}Magnetization switching as a thermal effect:
N\'eel-Brown model}
We have seen that the Stoner-Wohlfarth
model accounts for the hysteretic rotation of the particle's magnetization over
the potential barrier under the influence of a field applied in an arbitrary
direction, at zero temperature. We also considered the effect of temperature
on the Stoner-Wohlfarth model but at quasi-equilibrium.
Now, we come to deal with the thermoactivated switching of the particle's
magnetic moment, a process that occurs at short-time scales.
At nonzero temperatures the magnetization vector of the particle can
surmount the energy barrier due to thermal fluctuations as argued by N\'eel
\cite{nee49ancras}. This effect is particularly pronounced for small particles with
lower values of the potential barrier (\ref{barrier_height}).
Indeed, the magnetization vector of the particle shuttles between the two energy
minima and the characteristic time for this thermoactivated rotation of the
spin over the anisotropy barrier $\Delta{\mathcal H}$ is approximately given by the
celebrated Van't Hoff-Arrhenius law \cite{van1884}, \cite{arr1889}
\begin{equation}\label{arrhenius}
\tau = \tau_0\,e^{\Delta{\mathcal H}/k_BT},
\end{equation}
where $\tau_0$ is usually taken as temperature and field independent and on the
order of $10^{-10}-10^{-12}$s. $\tau_0$ is not necessarily the same
for different ferromagnetic materials.
It can indeed be assumed as constant only if the magnetization vector is
always in one of the energy minima, which happens only if the the minima have
zero width, or equivalently if the barrier is infinitely high.
However, in any realistic case, there is a finite probability that the
magnetization vector spends some time in the vicinity of either minimum, in
which case the prefactor need not be constant, and certainly depends on
temperature and field.
Indeed, in the case of cubic anisotropy the assumption of a
constant prefactor $\tau_0$ turns out to be a bad approximation \cite{aha96}.
For a given measuring time $\tau_m$ the system is at thermal equilibrium when
$\tau_m\gg \tau$, and in this case the particle is said to be {\it
superparamagnetic}, which corresponds to the temperature range
\begin{equation}\label{supa}
\ln(\tau_m/\tau_0) > \Delta{\mathcal H}/k_BT\geq 0.
\end{equation}
In fact, it was argued in \cite{garpal00acp} that due to the smallness of
$\tau_0$ the range of thermal equilibrium can extend down to temperatures at
which the energy-barrier height is much larger than thermal energy.
More precisely, for magnetic measurements with $\tau_m\sim 100$s, this range is
$0\leq \Delta{\mathcal H}/k_BT < 25$, which means that it is too restrictive to
argue that superparamagnetism occurs only when thermal energy is on the
order of the energy-barrier height.
In the case $\tau \gg \tau_m$ the particle's magnetization does not change during
the time of observation, and the particle exhibits stable ferromagnetism and is
said to be in a {\it blocked state}.
The temperature at which such a transition occurs, namely the temperature at
which the relaxation time $\tau$ is equal to the observation time $\tau_m$, is
called the {\it blocking temperature} and is denoted by $T_B$.
For nanoparticle assemblies with size distribution, the same temperature may
sometimes be above $T_B$ for some particles, and below it for the others.
Such systems may thus behave as superparamagnetic for some high values of
the temperature $T$, as ferromagnetic at low values of $T$, and as a mixture of
both at intermediate $T$.
For such a system a different characteristic temperature is used, this is
denoted by $T_{\max}$, and is roughly given, at least in the dilute case, by an
average of all $T_B$'s.
One should not forget however that the scale of $100$s depends on the
experimental apparatus. Indeed, in M\"ossbauer effect measurements, the
experimental time is the time of Larmor precession, which is on the order of
$10^{-8}$s, while in neutron scattering experiments it is on the order of
$10^{-12}$s.
In addition, the time scale can be completely different in different areas of
applications. For example, in magnetic recording, in order to keep the data
stored on a magnetic tape for years, one should have $\tau \gg 10^8$s, and in
rock magnetism the magnetization decays (or relaxes) within geological times
which may be millions of years.

Owing to these important practical applications and many others, the relaxation
time of the particle's magnetization is a very important and fundamental
physical quantity that deserves extensive and rigorous investigation.
This actually started, in this context, with the work of Kramers
on transition-state method \cite{kra40physa}.
Kramers showed, by using the theory of the Brownian motion, how the
prefactor of the reaction rate (inverse of relaxation time), as a function of
the damping parameter and the shape of the potential well, could be calculated
from the underlying probability-density diffusion equation in phase space, which
for Brownian motion is the Fokker-Planck equation (FPE). He obtained, by
linearizing the FPE about the potential barrier, explicit results for the escape
rate for intermediate-to-high (IHD) values of the damping parameter and also
for very small values of that parameter.
Subsequently, a number of authors \cite {mesmel86jcp}, \cite{butetal83prb}
showed how this approach could be extended to give formulae for the reaction
rate which are valid for all values of the damping parameter. These calculations
have been reviewed by H\"{a}nggi et al. \cite{hanetal90rmp}.

Kramers theory, originally developed for mechanical particles, was first
adapted to the thermal rotational motion of the magnetic moment by Brown
\cite{bro63pr} in order to improve upon N\'eel's concept of the
superparamagnetic relaxation process, which implicitly assumes discrete
orientations of the magnetic moment and which does not take into account the
gyromagnetic nature of the system.
Brown in his first explicit calculation \cite{bro63pr} of the escape rate
confined himself to axially symmetric (functions of the latitude only)
potentials of the magneto-crystalline anisotropy so that the calculation of the
relaxation rate is governed [for potential-barrier height significantly greater
than $k_{B}T$] by the smallest non-vanishing eigenvalue $\lambda_1$ of the
Sturm-Liouville equation associated with the one-dimensional FPE.
Thus the rate obtained is valid for all values of the damping parameter. As a
consequence of this very particular result, the analogy with the Kramers theory
for mechanical particles only becomes fully apparent when an attempt is made to
treat non axially symmetric potentials of the magneto-crystalline anisotropy
that are functions of both the latitude and the longitude.
In this context, Brown \cite{bro63pr} succeeded in giving a formula for the
escape rate for magnetic moments of single-domain particles, in the
IHD damping limit, which is the analog of the Kramers
IHD formula for mechanical particles.
In his second 1979 calculation \cite{bro79ieee} Brown only considered this case.
Later Klik and Gunther \cite{kligun90jap}, by using the theory of
first-passage times, obtained a formula for the escape rate which is the
analog of the Kramers result for very low damping.
All these (asymptotic) formulae which apply to a general non-axially symmetric
potential, were calculated explicitly for the case of a uniform magnetic field
applied at an arbitrary angle to the anisotropy axis of the particle by
Geoghegan et al. \cite {geoetal97acp} and compare favorably with the exact
reaction rate given by the smallest non-vanishing eigenvalue of the FPE
\cite{cofetal95prb}, \cite{cofetal98prb}, \cite{cofetal98prl}, 
\cite{cof98acp} and with experiments on the relaxation time of single-domain
particles \cite{cofetal98prl}, \cite{cofetal98jpc}.

In both limiting cases considered by Brown the time dependence of the average
magnetization is a single exponential and the relaxation rate is given by the
eigenvalue $\lambda_1$ previously mentioned.
Subsequently, $\lambda_1$ was numerically calculated by Aharoni for arbitrary
values of $\Delta{\mathcal H}/k_BT$ without a magnetic field \cite{aha64pr} and in a
longitudinal magnetic field \cite{aha69pr}.
The correction terms for the high-barrier result for $\lambda_1$ were given by
Brown \cite{bro77physb}.
Various limiting cases and the corresponding expressions for $\lambda_1$ were
further investigated in
\cite{scuetal92prb}, \cite{besetal92prb}, \cite{aha92prb},
\cite{creetal94jap}, \cite{cofetal95prb}.
Let aside such limiting cases, the FPE for an assembly of
single-domain ferromagnetic particles cannot be solved analytically. The
magnetization relaxation curve in fact consists of an infinite number of
exponentials.
In this case, it was shown in Refs.~\cite{garetal90tmp}, \cite{gar96prb2},
\cite{gar99epl} that it is more convenient to introduce the so-called {\it
integral relaxation time} $\tau_{int}$, defined as the area under the relaxation
curve after a sudden infinitesimal change of the magnetic field.
$\tau_{int}$ depends on all eigenvalues $\lambda_k, k=1,2,\ldots$ and therefore
contains more information than can be rendered by the first eigenvalue
$\lambda_1$, and more importantly, it can be measured experimentally.
Moreover, it turns out that, unlike $\lambda_1$, $\tau_{int}$ can be calculated
analytically for uniaxial particles in the longitudinal magnetic field for the
arbitrary values of the damping parameter, and $\tau_{int}^{-1}$ recovers the
analytical results of Brown for $\lambda_1$ in the asymptotic regions.
It was shown in \cite{gar96prb2} that the drastic deviation of $\tau_{int}^{-1}$
from $\lambda_1$ at low temperatures, starting from some critical value of the
field, is a consequence of the depletion of the upper potential well.
In these conditions, the integral relaxation time $\tau_{int}$ consists of two
competing contributions corresponding to the overbarrier and intrawell
relaxation processes.

After the above survey, we now give some of the most commonly used
expressions for the relaxation time, together with some cautionary remarks.
As indicated above, Brown \cite{bro63pr} at first derived a formula for
$\lambda_1$, for an arbitrary {\it axially symmetric} bistable potential of
Fig.~\ref{sw_landscape} with minima at $\theta =(0,\pi )$ separated by a
maximum at $\theta_m=\arccos(-h)$, which when applied to Eq.~(\ref{osp2}) for
${\bf h\Vert e,}$ i.e. a magnetic field parallel to the easy axis, leads to the
form given by Aharoni \cite {aha69pr},
\begin{equation}\label{lambda1_aharoni}
\lambda_{1}\simeq
\frac{2\alpha\gamma K^{3/2}}{m}
\sqrt{\frac{\beta}{\pi}}(1-h^{2})\times \left[
(1+h)\;e^{-\beta K (1+h)^{2}}+(1-h)\;e^{-\beta K (1-h)^{2}}\right],
\end{equation}
where $\gamma$ is the gyromagnetic factor, $\alpha$ the damping parameter.
$0\leq h\leq 1,$ $h=1$ being the critical value at which the bistable nature of
the potential disappears [see section~\ref{sec:SW}].

In order to describe the non-axially symmetric asymptotic behavior, let us
denote by $\beta \Delta {\mathcal H}_-$ the smaller reduced barrier height of the
two constituting escape from the left or the right of a bistable potential
[see Eq.~(\ref{barrier_height}) for the axially-symmetric case].
Then for very low damping, i.e. for $\alpha\times \beta \Delta {\mathcal H}_-\ll
1$ (with of course the reduced barrier height $\beta \Delta{\mathcal H}_-\gg 1$,
depending on the size of the nanoparticle), we have \cite{bro63pr},
\cite{cof98acp} the following asymptotic expression for the inverse relaxation
time
\begin{eqnarray}\label{VLD}
\tau _{VLD}^{-1} &\simeq& \frac{\lambda}{2\tau_N}\\ \nonumber
&\simeq& \frac{\alpha}{2\pi }\left\{ \omega _{1}\times \beta
({\mathcal H}_{0}-{\mathcal H}_{1})e^{-\beta ({\mathcal H}_{0}-{\mathcal
H}_{1})}+\omega_{2}\times \beta ({\mathcal H}_{0}-{\mathcal H}_{2})e^{-\beta
({\mathcal H}_{0}- {\mathcal H}_{2})}\right\},
\end{eqnarray}
where $\alpha$ is the damping parameter and
\begin{equation}\label{tauN}
\tau_N = \frac{1}{\alpha}\frac{\beta m}{2\gamma},
\end{equation}
is the free-diffusion time, i.e., the characteristic time of diffusion in the
absence of the potential.
For the IHD limit, where $\alpha\times \beta \Delta
{\mathcal H}_{-}>1$ (again with the reduced barrier height $\beta \Delta
{\mathcal H}_{-}$ much greater than unity) we have \cite{cof98acp} the
asymptotic expression
\begin{equation}\label{IHD}
\tau_{IHD}^{-1}\simeq \frac{\Omega_0}{2\pi
\omega _{0}}\left\{\omega_1 e^{-\beta ({\mathcal H}_0-{\mathcal H}_1)} + \omega
_{2}e^{-\beta ({\mathcal H}_0-{\mathcal H}_2)}\right\} ,
\end{equation}
where
\begin{eqnarray*}
\omega_1^2 &=&\frac{\gamma^2}{m^2}c_{1}^{(1)}c_{2}^{(1)},\quad
\omega_2^2=\frac{\gamma^2}{m^2}c_{1}^{(2)}c_{2}^{(2)}, \quad
\omega_0^2 =-\frac{\gamma ^{2}}{m^2}c_1^{(0)}c_2^{(0)},\\
\Omega _{0} &=&\frac{\alpha\gamma}{2m}\left[
-c_{1}^{(0)}-c_{2}^{(0)}+\sqrt{(c_{2}^{(0)}-c_{1}^{(0)})^{2}-4
c_{1}^{(0)}c_{2}^{(0)}}\right].
\end{eqnarray*}
Here $\omega_1,\omega_2$ and $\omega_0$ are respectively the wells
and saddle angular frequencies associated with the bistable potential,
$\Omega_0$ is the damped saddle angular frequency and the $c_{j}^{(i)}$
are the coefficients of the truncated (at the second order in the direction
cosines) Taylor series expansion of the crystalline anisotropy and external
field potential at the wells of the bistable potential denoted by $1$ and $2$
and at the saddle point denoted by $0$.
A full discussion of the application
of these general formulae to the particular potential, which involves the
numerical solution of a quartic equation in order to determine the
$c_{j}^{(i)}$ with the exception of the particular field angle $\psi =\frac{
\pi }{4}$ or $\frac{\pi }{2}$, in Eq.(\ref{osp1}) is given in Refs.
\cite{geoetal97acp}, \cite{kennedythesis97}.

In Ref.~\cite{kacetal00jpcm}, we used either Eqs.~(\ref{lambda1_aharoni})
and (\ref{VLD}) or (\ref{lambda1_aharoni}) and (\ref {IHD}), and
solved the equation $\tau = \tau_m$ for the blocking temperature $T_B$, as a
function of the applied field, for an arbitrary angle $\psi$ between the easy
axis and the applied magnetic field.
In particular, for very small values of $\psi$ we used
Eq.~(\ref{lambda1_aharoni}), as the problem then becomes almost axially
symmetric and the arguments leading to Eqs.~(\ref{VLD}) and (\ref{IHD}) fail
\cite{bro63pr}, \cite{bro79ieee}, \cite{geoetal97acp}, and
appropriate connection formulae had to be used so that they may attain the
axially symmetric limit.
Application of the above asymptotic formulae to the calculation of the blocking
temperature $T_B$ (or $T_{\max}$) as a function of the applied field
appeared to recover the experimental observation, but this result turned
out to be spurious.
An explanation of this behavior follows (see also \cite{garetal99pre},
\cite{cofetal00jmmm}): in the non-axially symmetric IHD asymptote (\ref{IHD}),
which is formulated in terms of the Kramers escape rate, as the field tends to
zero, for high damping, the saddle angular frequency $\omega_0$ tends to zero.
Thus the saddle becomes infinitely wide and so the escape rate predicted by
Eq.~(\ref{IHD}) diverges [see similar discussion in Sect.~\ref{sec:hk_tsp}]
leading to an apparent rise in the blocking temperature until the field reaches
a value sufficiently high to allow the exponential in the Arrhenius terms to
take over.
When this occurs the blocking temperature decreases again in accordance with the
expected behavior. This is the field range where one would expect the
non-axially asymptote to work well.
In reality, as demonstrated by exact numerical calculations of the
smallest non vanishing eigenvalue of the Fokker-Planck matrix, the small
field behavior is not as predicted by the asymptotic behavior of Eq.~(\ref
{IHD}) (it is rather given by the axially-symmetric asymptote) because the
saddle is limited in size to $\omega_0$.
Thus the true escape rate cannot diverge, and the apparent discontinuity between
the axially-symmetric and non axially-symmetric results is spurious, leading to
an apparent maximum in $T_B(h)$.
In reality, the prefactor in Eq.~(\ref{IHD}) can
never overcome the exponential decrease embodied in the Arrhenius factor.
Garanin \cite{garetal99pre} has discovered bridging
formulae which provide continuity between the axially-symmetric
Eq.~(\ref{lambda1_aharoni}) and non-axially symmetric asymptotes leading to a
monotonic decrease of the blocking temperature with the field in accordance with
the numerical calculations of the lowest eigenvalue of the Fokker-Planck
equation and also in agreement with the fact that the field suppresses the
energy barrier upon which the blocking temperature decreases.

An illustration of this was given in Ref.~\cite{garetal99pre} [see also
\cite{cofetal00jmmm} for more detail] for the particular case of $\psi=\pi/2$,
that is a transverse field. If the escape rate is written in the form
$$
\tau^{-1}=\frac{\kappa}{\pi}A\exp(-\beta\Delta{\mathcal H})
$$
where $\kappa$ is the attempt frequency given by
\[
\kappa=\frac{2K\gamma}{m}\sqrt{1-h^{2}},
\]
then the factor $A,$ as predicted by the IHD formula, behaves as
$\alpha/\sqrt{h}$ for $h\ll 1,\alpha^{2},$ while for $h=0$, $A$ behaves as
$2\pi \alpha\sqrt{\sigma/\pi },$ which is obviously discontinuous.
So, a suitable interpolation formula is required. Such a formula (analogous to that
used in the WKBJ method \cite{fermi65}) is obtained by multiplying the factor
$A$ of the axially symmetric result by $e^{-\xi }I_{0}(\xi ),$ where $I_{0}(\xi
)$ is the modified Bessel function of the first kind, and $\xi =2\sigma h$ [see
\cite{garetal99pre}].
This interpolation formula, as is obvious from the large and small $\xi $
limits, automatically removes the undesirable $1/\sqrt{h}$ divergence of the
IHD formula and establishes continuity between the axially symmetric and
non-axially symmetric asymptotes for $\psi =\pi /2$, as dictated by the
exact solution.

It is apparent from the discussion of this section that the N\'{e}el-Brown
model for a single particle is unable to explain the maximum in $T_{\max}$,
observed in experiments, as a careful calculation of the
asymptotes demonstrates that they always predict a monotonic decrease in the
blocking temperature.
A possible explanation of the maximum of $T_{\max}$ as a function of the
applied field was given in \cite{kacetal00jpcm} where it was shown that the
non-linearity of the field dependence of the superparamagnetic contribution to
the assembly magnetization, volume distribution, and anisotropy are responsible
for this effect.

\section{\label{sec:hk_tsp}A first step beyond the N\'eel-Brown model: effect
of exchange interaction}
As discussed in the introduction, in order to calculate the relaxation time of a
nanoparticle in an inhomogeneous magnetic state induced by surface effects, one
has to resort to microscopic theories.
Then, one has to take account explicitly of microscopic interactions, such as
exchange and/or dipolar interactions, in addition to the magneto-crystalline and
surface anisotropy and applied magnetic field.
However, the problem associated with the generalization of Brown's theory to
include interactions and eventually surface effects is really involved and can
in general only be tackled numerically.
But before one can attack this problem, one needs to understand the
effect of, e.g., exchange interaction on the relaxation time of the minimal
system, i.e., a pair of atomic spins coupled via exchange interaction,
including of course the usual magneto-crystalline anisotropy and Zeeman terms.
Besides, this is the unique non-trivial step towards the above-mentioned
generalization, where analytical expressions can be obtained for the
relaxation time.
It was the purpose of Ref.~\cite{kac03epl} to solve this problem and to
compare with the N\'eel-Brown result in the one-spin approximation.
Accordingly, we studied the effect of exchange
coupling on the relaxation rate of a magnetic system of two spins within
Langer's approach.
We found a particular value of the exchange coupling, that is $j\equiv J/K =
j_c\equiv 1-h^2$, where $h\equiv H/2K$, at which the number of saddle points
changes.
For $j\leq j_c$ there are two saddle points and the reversal of the
spins proceeds in two steps: the first spin crosses one of the saddle points
into the anti-ferromagnetic state, and then the other spin follows across the
second saddle point, to end up in the ferromagnetic order of opposite
direction with respect to the initial one.
For $j > j_c$, the two spins are so strongly exchange-coupled that they cross a
single saddle point together.
For $j\gg j_c$ the N\'eel-Brown result for a single spin is
recovered.
As a byproduct, we also showed that Langer's quadratic approximation of the
potential energy at the saddle point fails when the exchange coupling 
assumes the critical value $j_c$ even in the IHD limit.

It is worth mentioning that the effect of
dipolar interactions on the relaxation time of magnetic systems has been
studied by a few authors.
In Ref.~\cite{doretal88jpc}, \cite{mortro94prl} this was done for
assemblies of nanoparticles where approximate expressions were obtained using
the (mean-field) barrier-height approach that is only valid in the limit of
infinite damping (no gyromagnetic effect).
In Ref.~\cite{jongar01epl} the authors used perturbation
theory and obtained an expression for the relaxation rate in the case of weakly
interacting superparamagnets with various anisotropy orientations and in the
absence of external field.
Finally, the authors of Ref.~\cite{lybcha93jap} dealt with a pair of
coupled dipoles using the (numerical) Langevin approach and found some results
similar to ours [see below].

In this section we will first give a brief account of Langer's approach
\cite{lan68prl69ap} (see also \cite{bra94jap94prb} for
uniform (one-spin problem) and non-uniform magnetization and
\cite{cofgarmcc01acp} for comparison with Kramers' theory).
\subsection{\label{Lang_app}Switching rate in Langer's approach}
Within this approach the problem of calculating the relaxation time for a
multidimensional process is reduced to solving a steady-state equation in the
immediate neighborhood of the saddle point that the system crosses as it goes
from a metastable state to another state of greater stability.
The basic idea \cite{lan68prl69ap} is that ``{\it ... one imagines setting up a
steady-state situation by continuously replenishing the metastable
state at a rate equal to the rate at which it is leaking across the
activation-energy barrier. By identifying the current
flowing over the barrier with the desired decay rate, one avoids having to
solve the complete time-dependent problem...}'', especially in the
multidimensional case where this problem is too difficult to solve.
However, this approach is only valid in the limit of intermediate-to-high
damping because of the assumption, inherent to this approach, that the potential
energy in the vicinity of the saddle point may be approximated by its
second-order Taylor expansion (see below).
The result for small damping fails because the region of deviation from the
Maxwell-Boltzmann distribution, set up in the well, extends far beyond the
narrow region at the top of the barrier in which the potential may be replaced
by its quadratic approximation.
Therefore, in Langer's approach, one concentrates on the
distribution function $\rho(\{\eta\},t)$ as the probability that the
system is found in the configuration $\{\eta\}$ at time $t$. The time
evolution of $\rho$ is governed by the following equation
\begin{equation}\label{dyn_rho}
\frac{\partial\rho}{\partial t} =
\left (\frac{\partial\rho}{\partial t}\right)_{dyn.}
+ \left (\frac{\partial\rho}{\partial t}\right)_{fluct.}
\end{equation}
where the first term accounts for the ``internal'' dynamics of the system
described by
\begin{equation}\label{dyn_eq}
\frac{\partial\eta_i}{\partial t} = \sum_j A_{ij}\,\frac{\partial{\mathcal
H}}{\partial\eta_j},
\end{equation}
where $A$ is a fully anti-symmetric matrix. For a magnetic system, as
in our case, the analog of Eq.~(\ref{dyn_eq}) is given by the
Landau-Lifshitz equations (\ref{LinLLE}).
The second term in (\ref{dyn_rho}) accounts for the dynamics of the
system driven by its interaction with the heat bath, and is given by
\cite{lan68prl69ap}
\begin{equation}\label{fluct}
\left (\frac{\partial\rho}{\partial t}\right)_{fluct.} = \sum_i
\Gamma_i\frac{\partial}{\partial\eta_i}\left (\beta
\frac{\partial{\mathcal H}}{\partial\eta_i}
+ \frac{\partial\rho}{\partial\eta_i}\right).
\end{equation}
where $\beta=1/k_BT$ and the constant $\Gamma_i$ characterizes the
variation rate of $\eta_i(t)$.
Combining Eqs.~(\ref{dyn_eq}) and (\ref{fluct}) in Eq.~(\ref{dyn_rho})
leads to the Fokker-Planck equation
\begin{equation}\label{FPE_langer}
\partial_t\rho +\sum_i\partial_{\eta_i}J_{i}=0,
\end{equation}
which is a continuity equation with the probability current
\begin{equation}\label{Ji}
J_{i}=-\sum _{j}M_{ij}\left( \partial _{\eta_j}H+\frac{1}{\beta
}\partial_{\eta_j}\right) \rho,
\end{equation}
and where the matrix $M$ reads
\begin{equation}\label{matrix_M}
M_{ij}\equiv \beta \Gamma _{i}\delta_{ij}-A_{ij}.
\end{equation}
In order to calculate the nucleation rate, one must solve
Eq.~(\ref{FPE_langer}). A particular solution is obtained at equilibrium and is
given by the Maxwell-Boltzmann distribution
\begin{equation}\label{rho_eq}
\rho _{eq} = \exp (-\beta {\mathcal H}\left\{\eta\right\})/Z_0,
\end{equation}
which corresponds to zero current, $J_{i}=0$. However,
what is really needed is to obtain a finite probability current flowing across
the saddle point $\eta^{(s)}$. As stated above, considering a
steady-state so that the rate at which the metastable state is
replenished is equal to the rate of leak across the activation-energy
barrier, the problem then consists in finding this steady-state as a
solution of Eq.~(\ref{FPE_langer}) in the immediate neighborhood of the
saddle point.
To obtain this solution, it is more convenient to work in the frame of
canonical coordinates with their origin at $\eta^{(s)}$, defined by
\begin{equation}\label{xi's}
\xi_n=\sum_i\mathcal{D}_{ni}(\eta_i-\eta_i^{(s)}),
\end{equation}
where the transformation matrix $\mathcal{D}$ diagonalizes the energy
Hessian, so that the energy can be rewritten near $\left\{ \xi
\right\} =\left\{ 0\right\}$ as
\begin{equation}\label{E_in_xi's}
E_{s}\simeq E^{(0)}_{s}+\frac{1}{2}\sum_{n=1}^N\lambda_{n}\xi _{n}^{2}.
\end{equation}
The $\lambda_n$'s are the eigenvalues of the energy Hessian. By the
definition of $\eta^{(s)}$, one of the $\lambda$'s, denoted in the
sequel by $\lambda_{-}$, is negative; that is, the energy at the
saddle point diminishes on either
side of the surface $\xi_-=0$. Moreover, we expect that $m<N$ eigenvalues
vanish, which corresponds to the fact that the saddle point does not possess all
of the symmetries of ${\mathcal H}\left\{\eta\right\}$.
The unbroken symmetries produce Goldstone modes \footnote{A spontaneous breaking
of a continuous symmetry entails the existence of a massless mode, that is a
zero-energy fluctuation, called the Goldstone mode: This is the well-known
Goldstone theorem~\cite{goldstonetheorem}.} that must be eliminated from the
nucleation rate as they cause the latter to diverge.
Consequently, the saddle point is actually a bounded, finite-dimensional
subspace of the $\eta$-space. The volume of this subspace will be
denoted below by ${\mathcal V}$. For the single-spin problem, this is just
the space spanned by rotations around the angle $\varphi$, while
$\cos\theta^{(s)}=-h$, and its volume is given by ${\mathcal V}_s=2\pi
\sin\theta^{(s)}=2\pi\sqrt{1-h^2}$. \\
Next, one diagonalizes the transition matrix that results from the
dynamical equation (\ref{dyn_eq}) linearized at the saddle point, that
is one has to solve the eigenvalue problem:
\begin{equation}
\label{EVpb}
\lambda_n\sum_m^{'}\tilde{M}_{nm}\,U_{m}=\kappa U_n,
\end{equation}
where the prime on the summation indicates that we omit all n's for which
$\lambda_n=0$, and
\begin{equation}\label{Mtilde}
\tilde{M}_{nm}\equiv
\sum_{i,j}\mathcal{D}_{ni}(-A_{ij})\mathcal{D}_{mj}=-\tilde{M}_{nm}.
\end{equation}
As was argued by Langer \cite{lan68prl69ap}, one, and only one, of
the eigenvalues $\kappa$ must be negative. Indeed, if the saddle point is to
describe the nucleating fluctuation, there must be exactly one direction of
motion away from the saddle point $\left\{ \xi \right\} =\left\{ 0\right\}$ in
which the solution
\begin{equation}\label{sol_xi_n}
\left\langle \xi_m\right\rangle = X_n\, e^{-\kappa t},
\end{equation}
of the equation of motion of the $\xi_n$ modes
\begin{equation}
\label{EM_xi_n}
\frac{d\left\langle\xi_n\right\rangle }{dt}=-\sum_m
\tilde{M}_{nm}\, \lambda_m\, \left\langle \xi_m\right\rangle,
\end{equation}
is unstable.

Finally, $Z_{0}$ in Eq.~(\ref{rho_eq}) is a normalization factor defined by
the condition that the equilibrium probability density (\ref{rho_eq}) is
normalized at the metastable state, i.e.,
$$
\frac{1}{Z_{0}}\int_{\eta^{(m)}}\, \mathcal{D}\eta\, \exp (-\beta
{\mathcal H}^{(m)}\{\eta\})=1,
$$
which leads to
\begin{equation}\label{Z_0}
Z_{0}\simeq e^{-\beta E^{(0)}_{m}}\prod _{l=1}^{4}\left( \frac{2\pi
}{\beta \lambda_{l}^{(m)}}\right)^{1/2},
\end{equation}
where $\lambda _{l}^{m}$ are the eigenvalues of the energy Hessian
at the metastable state and $E^{(0)}_{m}$ its energy.

Taking all this into consideration, Langer has given a formula for the
switching rate $\Gamma$ out of the metastable state which reads,
\begin{equation}\label{Gamma}
\Gamma =\frac{{\mathcal V}}{2\pi}
\sqrt{\frac{\beta}{2\pi}}
\frac{\left|\kappa\right|}{\sqrt{\left|\lambda_-\right|}}
\left(\prod_{l}\sqrt{\lambda_{l}^{m}}\right)
\left(\prod_{n}^{''}\frac{1}{\sqrt{\lambda _{n}}}\right)
\times e^{-\beta \Delta E_{0}},
\end{equation}
where $\Delta E_{0}\equiv E_{0}^{(s)} - E_{0}^{(m)}$ is the
barrier energy between the saddle point and the metastable state, and
the double prime on the product symbol indicates that we omit the
negative eigenvalue $\lambda_-$ and all $n$ for which $\lambda_n=0$.

The whole procedure can be summarized as follows:
From a static study of the energy, one identifies the metastable states
and saddle points. Then at a given saddle point one expands the energy to second
order, calculates the Hessian and its eigenvalues.
The vanishing eigenvalues corresponding to the Goldstone
modes, associated with the unbroken symmetries at the saddle point,
are eliminated from the final expression of the relaxation
rate.
The volume ${\mathcal V}$ of the saddle-point subspace is also
calculated. Then, one solves the eigenvalue problem (\ref{EVpb}) and
obtains the negative eigenvalue $\kappa$.
Finally, one calculates the eigenvalues $\lambda_l^{m}$ of the Hessian of energy
expanded at the metastable state.

A quite similar approach has been used by D.A. Garanin \cite{gar91jp1} to
calculate the thermo-activation rate of charged particles in a magnetic field.
Here, the Fokker-Planck equation is written in terms of the 6 variables, 3
spatial coordinates and 3 momenta,  and then linearized near the saddle point.
The problem then amounts to computing the flux and normalization in the wells
rendering a compact expression for the rate in terms of the frequencies in the
wells and at the saddle point.

\subsection{\label{tsp_langer}Relaxation rate for the two-spin problem within
Langer's approach}
We consider a system of two (classical) exchange-coupled spins with the
Hamiltonian
\begin{eqnarray}
&&{\mathcal H} =-\frac{j}{2}\vec{s}_{1}\cdot\vec{s}_{2}-\frac{1}{2}\left[
(\vec{s}_{1}\cdot\vec{e}_{1})^{2}+(\vec{s}_{2}\cdot\vec{e}_{2})^{2}\right]
-\vec{h}\cdot(\vec{s}_{1}+\vec{s}_{2}), \nonumber \\
&&j\equiv J/K,\, h\equiv H/2K  \label{HTS}
\end{eqnarray}
where $J>0$ is the exchange coupling, $K>0$ the
anisotropy constant, $H$ the applied magnetic field, and $h$
the reduced field, i.e., $0\leq h < 1$. $\vec{e}_i$ are uniaxial
anisotropy unit vectors.
%, see Fig.~\ref{tsp}.
%
%%%%%%%%%%%%%%%%%%%%%%%%%%%%%%%%%%%%%%%%%%%%%%%%%%%%%%%%%%%%%%%%%%%%%%
%\begin{figure}[h!]
%\includegraphics[width=5cm]{tsp.eps}
%\caption{\label{tsp}System of two spins with easy axes along the field
%applied along the $z$ direction.}
%\end{figure}
%%%%%%%%%%%%%%%%%%%%%%%%%%%%%%%%%%%%%%%%%%%%%%%%%%%%%%%%%%%%%%%%%%%%%%
%
We use the dimensionless units, i.e., $\left[{\mathcal H}\right]=2K$ for energy
and $\left[t\right]=1/(2\gamma K)$ for time.
Here we restrict ourselves to the case
$\vec{h}\parallel\vec{e}_i, i=1,2$, with the same anisotropy constant.
Owing to the symmetry of this system with respect to rotations around the easy
axis, the number of variables reduces to three, $\theta_1, \theta_2,
\varphi\equiv\varphi_1-\varphi_2$.

\begin{figure}[t!]
\includegraphics[angle=-90, width=16cm]{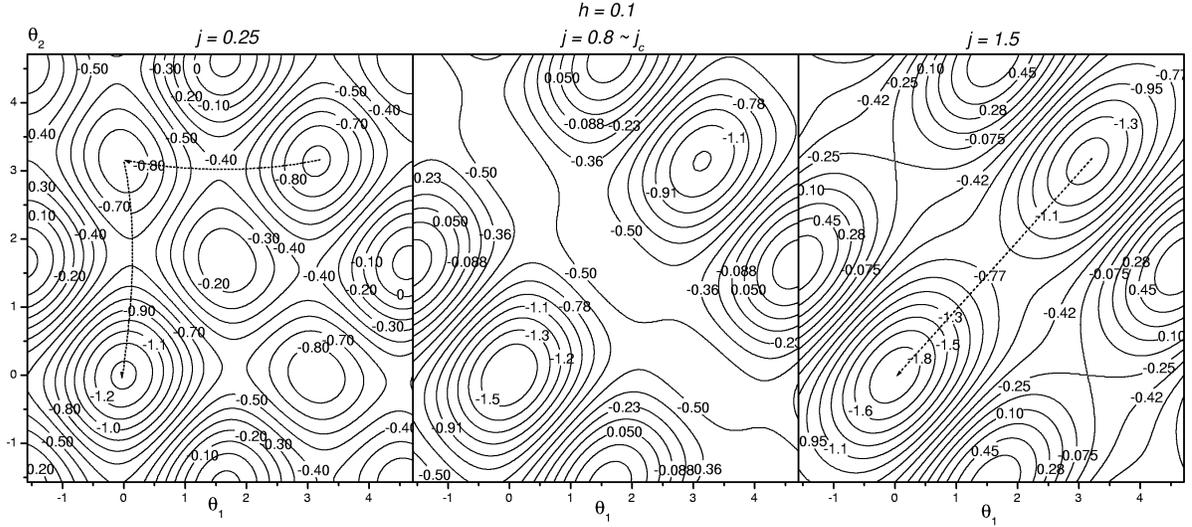}
\vspace{-3cm}
\caption{\label{enscape}
Energyscape from Eq.~(\ref{HTS}) ($\varphi=0$) for $j=0.25, 0.8, 1.5$ and
$h=0.1$. The arrows indicate the switching paths.}
\end{figure}

Now we apply Langer's approach to the energy (\ref{HTS}) and study the
relaxation rate as a function of the exchange coupling $j$.
We first analyze the energyscape in Fig.~\ref{enscape}.
%
%%%%%%%%%%%%%%%%%%%%%%%%%%%%%%%%%%%%%%%%%%%%%%%%%%%%%%%%%%%%%%%%%%%%%%%%%%%%%
%%%%%%%%%%%%%%%%%%%%%%%%%%%%%%%%%%%%%%%%%%%%%%%%%%%%%%%%%%%%%%%%%%%%%%%%%%%%%
%
The absolute minimum of the energy (\ref{HTS}) corresponds to the
ferromagnetic ordering of the two spins along the easy axis,
\begin{equation}
(0,0,\varphi);\quad e_{ss}^{(0)}=-\frac{j}{2}-2h-1,  \label{sfm}
\end{equation}
where henceforth $e^{(0)}$ denotes the energy of the state.
One metastable state corresponds to a ferromagnetic ordering opposite to
the field
\begin{equation}
(\pi,\pi;\varphi);\quad e_{m_1}^{(0)}=-
\frac{j}{2}+2h-1.  \label{mfm}
\end{equation}
There is also the metastable state of anti-ferromagnetic ordering,
\begin{equation}\label{afm}
(0,\pi;\varphi)\,\mbox{or}\,(\pi,0;\varphi);\quad
e_{m_2}^{(0)}=\frac{j}{2}-1.
\end{equation}
As to saddle points, we find that their number and loci crucially depend on
the exchange coupling constant $j$. More precisely, for $j
>j_c\equiv 1-h^2$, there is a single saddle point given by
\begin{equation}
(\cos\theta_1=\cos\theta_2 = -h; \varphi);\quad
e_{s}^{(0)}=-\frac{j}{2}+h^{2},  \label{SP0}
\end{equation}
whereas for $j < j_c$ there are two saddle points given by
\begin{eqnarray}\label{SP21}
\cos\theta_{1,2}^{\varepsilon}= \frac{1}{2}\left( -h-a^{\varepsilon}\pm
\sqrt{\Delta^{\varepsilon}}\right)\equiv X^{\varepsilon}_{\pm}; \,\, \varphi, &  &
\end{eqnarray}
where $\varepsilon=\pm$ and
\begin{eqnarray*}
&& a^{\varepsilon} = \varepsilon\sqrt{1-j} \quad b^{\varepsilon} =
\varepsilon\sqrt{1+\frac{(j/2)^2}{1-j}} \nonumber \\
&& \Delta^{\varepsilon}=(h+a^{\varepsilon})^2+2j-4b^{\varepsilon} h,
\end{eqnarray*}
%
%where $\varepsilon=\pm, a^{\varepsilon} = \varepsilon\sqrt{1-j},
%b^{\varepsilon} = \varepsilon\sqrt{1+(j/2)^2/(1-j)}$, and
%$\Delta^{\varepsilon}=(h+a^{\varepsilon})^2+2j-4b^{\varepsilon}h,$
%
with energy,
\begin{eqnarray}
&& e_{\varepsilon}^{(0)}=-\frac{j}{2}\sqrt{(1+b^{\varepsilon}
h-\frac{j}{2})^2-(a^{\varepsilon}+ h)^2} - \frac{j}{2}(1+b^{\varepsilon}
h-\frac{j}{2}) \nonumber \\
&& + \frac{1}{2}(h^2-(a^{\varepsilon})^2+2b^{\varepsilon} h).
\label{enj<jc}
\end{eqnarray}
At $j=j_c$ the saddle point (\ref{SP21}) with
$\varepsilon=+$ merges with the saddle point (\ref{SP0}), while that with
$\varepsilon=-$ merges with the metastable state (\ref{afm}),
see Fig.~\ref{enscape} central panel.

\begin{figure}[h!]
\includegraphics[angle=-90,width=12cm]{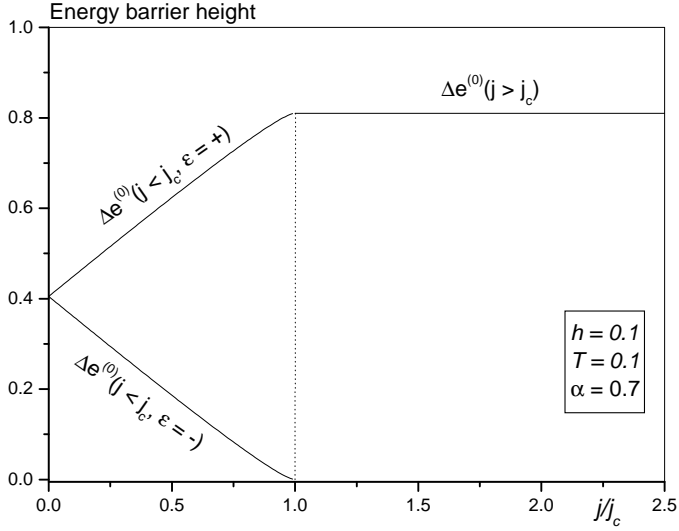}
\caption{\label{BH}
Energy barrier as a function of $j$.
}
\end{figure}

Starting with both spins aligned in the metastable
state (\ref{mfm}) with $\theta_{1,2}=\pi$, if $j < j_c$ one of the
two spins crosses the saddle point (\ref{SP21})($\varepsilon=+$) into the state
(\ref{afm}) by reversing its direction.
Then the second spin follows through the second saddle point
(\ref{SP21})($\varepsilon=-$), of lower energy due to the
exchange coupling (see Fig.~\ref{BH}), ending up in the stable
state (\ref{sfm}).
In Fig.~\ref{enscape} (left) the path is indicated by a pair of curved arrows.
There are actually two such paths corresponding to the two-fold symmetry of
the problem owing to the full identity of the two spins.
Note that when the first spin starts to switch and arrives at
$\theta_1\sim\pi/2$, the second spin has $\theta_2\lesssim\pi$ (hence the
curved arrows in Fig.~\ref{enscape}), which suggests that in the switching
process of the first spin, the position of the second spin
undergoes some fluctuations creating a small transverse field, and when
$\theta_1=0$ the second spin goes back to the position $\theta_2=\pi$
before it proceeds to switch in turn.
The successive switching of the two spins through the corresponding saddle
points is a sequential two-step process
\footnote{\label{footnoteChLyb}Similarly, it was found in
Ref.~\cite{lybcha93jap} that the reversal of the two dipoles considered
is a two-stage process with an intermediary metastable antiparallel state.}, so
the relaxation rates for $j < j_c$ add up inverse-wise.
In the case $j > j_c$ the two spins cross the
unique saddle point (\ref{SP0}) to go from the metastable state (\ref{mfm}) into
the stable one (\ref{sfm}) in a single step, see Fig.~\ref{enscape} (right)
where the path is indicated by a single straight arrow.
There is the symmetry $\pm\theta^{(s)}$,
which leads to a factor of 2 in the relaxation rate.
Therefore, if we denote by $\Gamma_{j\leq j_c}^+$, $\Gamma_{j\leq j_c}^-$,
and $\Gamma_{j \geq j_c}$ the respective relaxation rates, the relaxation rate
of the two-spin system can then be written as
\begin{equation}\label{Gamma_tsp}
\Gamma = \left\{
\begin{array}{ll}
2\Gamma_{j\leq j_c}^+ \Gamma_{j\leq j_c}^-/(\Gamma_{j\leq
j_c}^+ + \Gamma_{j\leq j_c}^-),  & \mbox{if $j \leq j_c$} \\ \\
2\Gamma_{j \geq j_c}, & \mbox{if $j \geq j_c$}.
\end{array}
\right.
\end{equation}
%
%%%%%%%%%%%%%%%%%%%%%%%%%%%%%%%%%%%%%%%%%%%%%%%%%%%%%%%%%%%%%%%%%%%%%%%%%%%%%
%%%%%%%%%%%%%%%%%%%%%%%%%%%%%%%%%%%%%%%%%%%%%%%%%%%%%%%%%%%%%%%%%%%%%%%%%%%%%
%

In Langer's approach the $\Gamma$'s in Eq.~(\ref{Gamma_tsp}) are obtained from
the steady-state Fokker-Planck equation (SSFPE) linearized around each saddle
point, using the fluctuating variables $\eta_i=(t_i,p_i)$ where
$t_i\equiv\theta_i-\theta_i^s,\,p_i\equiv\varphi_i-\varphi^s_i , \, i=1,2$, or
more adequately the ``canonical" variables
$\psi_n=(\xi_{\pm},\zeta_{\pm})$ with $\xi_{\pm}=(t_1\pm
t_2)/\sqrt{2},\,\zeta_{\pm}=(p_1\pm p_2)/\sqrt{2}$,
in which the energy Hessian is diagonal.

The deterministic dynamic of the system is governed by the Landau-Lifshitz
equations, which upon linearization near the saddle point, read
\begin{equation}\label{LinLLE}
\left\{
\begin{array}{ll}
\partial_t\,t_i=-\partial_{p_i}{\mathcal H}_2-\alpha \partial_{t_i}{\mathcal
H}_2\\
\partial_t\,p_i=-\alpha\partial_{p_i}{\mathcal H}_2 + \partial_{t_i}{\mathcal
H}_2,\quad i=1,2
\end{array}
\right.
\end{equation}
where ${\mathcal H}_{2}$ is the quadratic approximation of the energy (\ref{HTS})
near the saddle point and $\alpha$ is the damping parameter.
In matrix form Eqs.~(\ref{LinLLE}) become
\begin{equation}\label{LLE_langer}
\partial _{t}\eta _{i}=\sum _{j}M_{ij}\, \partial _{\eta
_{j}}{\mathcal H}_{2},
\end{equation}
where $M$ is the dynamic matrix containing the precessional and
dissipative parts.
Rather than investigating the stochastic trajectories
$\eta_{i}(t)$ that arise by adding a noise term in Eq.~(\ref{LLE_langer}), Langer
concentrates on the distribution function $\rho(\{\eta\},t)$ as the probability
that the system is found in the configuration $\{\eta\}$ at time $t$. The time
evolution of $\rho$ is governed by the Fokker-Planck equation
(\ref{FPE_langer}), where now the current $J_i$ is given by Eq.~(\ref{Ji}) with
the Hamiltonian ${\mathcal H}$ replaced by its quadratic approximation ${\mathcal H}_2$.

Instead of solving the time-dependent equation (\ref{FPE_langer}),
Langer solves the SSFPE $\partial_t\rho = 0$ near the saddle point.
The steady-state is realized by imposing the boundary conditions:
$\rho\simeq\rho_{eq}$ near the metastable state and $\rho\simeq 0$ beyond the
saddle point.
Then, the problem of calculating the escape rate boils down to the calculation
of the total current by integrating the probability current (\ref{Ji}) over a
surface through the saddle point.
After performing all these steps, Langer arrives at his famous expression
(\ref{Gamma}) for the relaxation rate which is valid in the IHD limit.
The general calculations can be found in greater details in \cite{lan68prl69ap},
or applied to magnetic particles in \cite{bra94jap94prb}, see also the review
article \cite{cofgarmcc01acp}.

However, since the escape rate is simply given by the ratio of the total current
through the saddle point to the number of particles in the metastable state, it
turns out that Langer's result for the escape rate can be retrieved by only
computing the energy-Hessian eigenvalues near the saddle points and metastable
states, from which one then infers the partition function $\tilde{Z_s}$ of the
system restricted to the region around the saddle point where the energy-Hessian
negative eigenvalue is (formally)\footnote{See Ref.~\cite{lan68prl69ap} for a
rigorous derivation} taken with absolute value, and the partition function $Z_m$
of the region around the metastable state.
When computing these partition functions, one has to identify and take care of
the Goldstone modes associated with the unbroken global symmetries of the
different states.
Finally, one computes the unique negative eigenvalue $\kappa$ of the SSFPE
corresponding to the unstable mode at the saddle point.
More precisely, $\kappa$ is given by the
negative eigenvalue of the dynamic matrix
$\tilde{M}_{mn}=-\lambda_n({\mathcal D}M{\mathcal D}^T)_{mn}$, where the $\lambda_n$'s are
the eigenvalues of the Hessian at the saddle point and ${\mathcal D}$ is the transformation
matrix from $\eta_i$ to $\psi_n$ [see Eq.~(\ref{xi's})].

Consequently, Langer's final expression for the escape rate is rewritten
in the following somewhat more practical form
\begin{equation}\label{Gammapr}
\Gamma=\frac{\left|\kappa\right|}{2\pi}\frac{\tilde{Z_s}}{Z_m},
\end{equation}
where $\left|\kappa\right|$ is the attempt frequency which contains the
damping parameter $\alpha$.

\subsection{Relaxation rate of the two-spin system}
Now, we give the different relaxation rates for $j > j_c, j < j_c$, and
$j\simeq j_c$.

\subsubsection{Relaxation rate for $j > j_c$}
The quadratic expansion of the energy (\ref{HTS}) at the saddle point
(\ref{SP0}) reads,
\begin{equation}\label{QEj>j_c}
H^{(2)}_s=e_s^{(0)}-
\frac{j_c}{2}\xi_{+}^2+\frac{j-j_c}{2}\xi_{-}^2+\frac{jj_c}{2}
\zeta_{-}^2,
\end{equation}
with zero eigenvalue for the $\zeta_+$ mode, that is the Goldstone
mode associated with the rotation around the easy axis. $e_s^{(0)}$ is
given in Eq.~(\ref{SP0}).
The negative eigenvalue of the SSFPE, corresponding to the unstable mode,
is $\kappa =-\alpha j_c$.

The partition function around the saddle point $Z_{s}=\int d\Omega_1
d\Omega_2 e^{-\beta H_{s}^{(2)}},$
where $d\Omega_i=\sin\theta_i d\theta_i d\varphi_i,\, i=1,2$ is calculated
by changing to the variables $\xi_{\pm},\zeta_{\pm}$, setting
$\sin\theta_i\simeq \sqrt{j_c}$ (see Eq. (\ref{SP0})) in the
integration measure, and substituting $2\pi$ for the integral over $\zeta_+$,
and finally computing the Gaussian integrals. Hence,
\begin{equation}
\label{Z_sp}
\tilde{Z}_{s}=2\pi \left(
\frac{2\pi}{\beta}\right)^{3/2}\frac{1}{j\sqrt{1-j_c/j}}e^{-\beta
e^{(0)}_{s}},
\end{equation}
where we have formally replaced the negative eigenvalue $(-j_c)$ by
its absolute value.
Next, the partition function $Z_m$ at the (well defined) metastable state
(\ref{mfm}) is computed by expanding the energy up to $2^{nd}$
order, leading to
\begin{equation}\label{Z_m}
Z_{m}=e^{-\beta e^{(0)}_{m1}}\left( \frac{2\pi }{\beta }\right) ^{2}\frac{1}{(1-h)(j+1-h)}.
\end{equation}

Finally, using Eq. (\ref{Gammapr}), the relaxation rate for $j>j_c$
reads, upon inserting the symmetry factor of $2$,
\begin{eqnarray}
&& \Gamma_{j>j_c} =2\alpha \sqrt{\frac{\beta }{2\pi
}}(1-h^{2})(1-h)\frac{1+(1-h)/j}{\sqrt{1-j_c/j}}\times e^{-\beta
\Delta e^{(0)}}, \nonumber \\
&& \Delta e^{(0)}=(1-h)^{2}. \label{Gamma_tsp_j>jc}
\end{eqnarray}

It is readily seen that for $j\rightarrow \infty$ (\ref{Gamma_tsp_j>jc}) tends
to the N\'eel-Brown result,
\begin{equation}\label{NB}
\Gamma_{j > j_c}\rightarrow 2\alpha
\sqrt{\frac{\beta }{2\pi }}(1-h^2)(1-h)\,\,e^{-\beta(1-h)^{2}},
\end{equation}
for the relaxation rate of one rigid pair of spins with a barrier height
twice that of one spin.
In fact, the result (\ref{NB}) coincides with the
N\'eel-Brown result in Eq.~(\ref{lambda1_aharoni}) upon reinstating the energy
and time units and remembering that the result (\ref{NB}) is for the
one-way barrier crossing.
Note that the convergence of (\ref{Gamma_tsp_j>jc}) to (\ref{NB}) is so slow
that $\Gamma_{j > j_c}$ remains above the N\'eel-Brown result and only
merges with it for $j\gtrsim 10$.
This confirms the fact that the one-spin approximation is only valid for
extremely large exchange interaction.

\subsubsection{Relaxation rate for $j < j_c$}
Here the attempt frequencies
$|\kappa^{\varepsilon}|$ (for $\varepsilon=\pm$) cannot be obtained in a
closed form and are thus computed numerically, this means that the present case
can only be dealt with semi-analytically.
On the other hand, following the same procedure as for $j > j_c$ we obtain
the relaxation rate for $\varepsilon=\pm$,
%
%%%%%%%%%%%%%%%%%%%%%%%%%%%%%
\begin{equation}\label{gamma_jleqjc}
\Gamma_{j < j_c}^{\varepsilon}=\sqrt{\frac{\beta}{\pi}}|\kappa^{\varepsilon}|
\sqrt{\frac{P^{\varepsilon}}{j}}\frac{N^{\varepsilon}}{\sqrt{R^{\varepsilon}_+
R^{\varepsilon}_- + jQ^{\varepsilon}(R^{\varepsilon}_+
+ R^{\varepsilon}_-)}} \, e^{-\beta\Delta e^{(0)}_{\varepsilon}},
\end{equation}
%%%%%%%%%%%%%%%%%%%%%%%%%%%%%
%
where $\Delta e^{(0)}_{\varepsilon}$ is the barrier height given by
the energy (\ref{enj<jc}) measured with respect to (\ref{mfm}) or
(\ref{afm}) for $\varepsilon=+,-$ respectively, and (see Eq.~(\ref{SP21})
et seq. for notation)
%
%%%%%%%%%%%%%%%%%%%%%%%%%%%%%
\begin{eqnarray*}
&& P^{\varepsilon} =
\sqrt{(1+b^{\varepsilon}h-j/2)^2-(a^{\varepsilon}+h)^2}, \\
&& Q^{\varepsilon}=b^{\varepsilon}h-j/2+P^{\varepsilon}, \quad
R^{\varepsilon}_{\pm} = -1+X^{\varepsilon}_{\pm}(2X^{\varepsilon}_{\pm} + h), \\
&& N^+ = (1-h)(j+1-h), \quad N^-=j_c-j.
\end{eqnarray*}
%%%%%%%%%%%%%%%%%%%%%%%%%%%%%
%
The limit of the relaxation rate (\ref{gamma_jleqjc})
when $j\longrightarrow 0$ is just the N\'eel-Brown result for one spin.
Indeed, the product of the last two factors in the prefactor tend to
$(1-h)/\sqrt{2}$, the attempt frequencies tend to $\alpha j_c=\alpha
(1-h^2)$, and the energy barriers $\Delta
e^{(0)}_\varepsilon\rightarrow (1-h)^2/2$.

Note that in the present regime of $j<j_c$ the large value of anisotropy
has not changed the temperature dependence of the individual relaxation
rates, i.e., $\Gamma_{j < j_c}^{\varepsilon}$, with $\varepsilon = \pm$.
This is due to the fact that $1/\sqrt{T}$ appears in the prefactor each
time there is a continuously degenerate class of saddle points
\cite{bra94jap94prb}, which is indeed the case for $j>j_c$ and $j<j_c$ with
$\varepsilon = +$ and $\varepsilon = -$.
However, anisotropy do affect the temperature dependence of the relaxation
rate of the two-spin system, since for $j < j_c$ there are two
saddle points bringing each a factor $1/\sqrt{T}$, see the first line in
Eq.~(\ref{Gamma_tsp}).

\subsubsection{The case of $j\simeq j_c$}
\begin{figure}[h!]
\includegraphics[angle=-90, width=13cm]{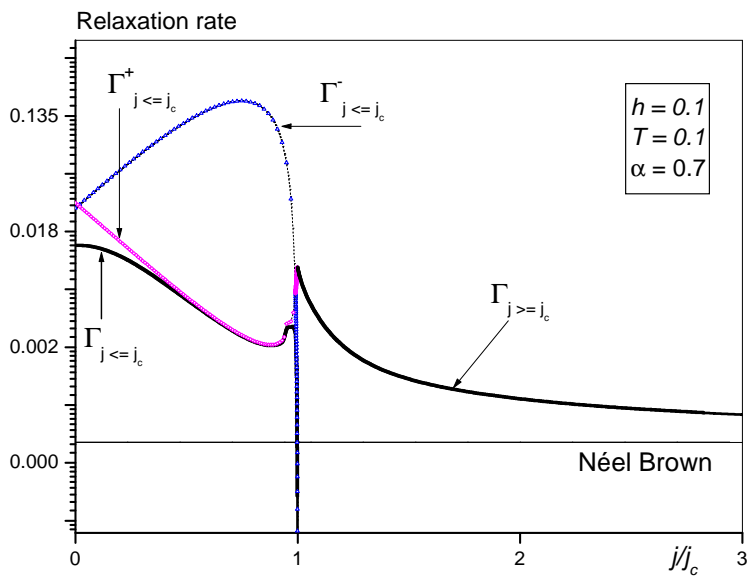}
\caption{\label{rate}
Relaxation rate (in logarithmic scale) of the two-spin system in the case
of intermediate-to-high damping. The horizontal line is the N\'eel-Brown
result with the barrier height taken twice that of a single spin.}
\end{figure}

When $j$ approaches $j_c$ either from above or
from below, more Hessian eigenvalues (in addition to $\lambda_{\zeta^+}$)
vanish, rendering the saddle point rather flat and thus leading to a divergent relaxation
rate.
Indeed, for $j\simeq j_c$ the
relaxation rate (\ref{Gamma_tsp_j>jc}) diverges, which clearly shows that
Langer's approach making use of a quadratic approximation for the energy at
the saddle point, e.g., Eq.~(\ref{QEj>j_c}), fails in this case.
The remedy is to push the energy expansion to the $6^{th}-$order in the variable
$\xi_-$ (since $\lambda_{\xi_-}=j-j_c\rightarrow 0$ as $j\rightarrow
j_c$), i.e.,
\begin{equation}\label{E6}
\delta e_s = e_s-e_s^{(0)}\simeq
\frac{\lambda_{\xi_-}}{2}\xi_-^2+\frac{c}{4}\xi_-^4+\frac{d}{6}\xi_-^6,
\end{equation}
where
%$c=(1-j-7h^2/4)/3<0, \,d=(j-1+31h^2/16)/30>0$.
%
$$
c = \frac{1-j-7h^2/4}{3}<0, \,\, d=\frac{j-1+31h^2/16}{30}>0.
$$
Then, the contribution $\sqrt{2\pi/\beta \lambda_{\xi_-}}$ of the mode $\xi_-$ to
the relaxation rate (\ref{Gamma_tsp_j>jc}) must be replaced
by$\int_{-\infty}^{\infty}d\xi_- e^{-\beta\delta e}$, upon which the
divergence of $\Gamma_{j>j_c}$ is cut out (see Fig.~\ref{rate}).
%
%%%%%%%%%%%%%%%%%%%%%%%%%%%%%%%%%%%%%%%%%%%%%%%%%%%%%%%%%%%%%%%%%%%%%%%%%%%%%
%%%%%%%%%%%%%%%%%%%%%%%%%%%%%%%%%%%%%%%%%%%%%%%%%%%%%%%%%%%%%%%%%%%%%%%%%%%%%

Similarly, when $j$ approaches $j_c$ from below,
for $\varepsilon=-$ the eigenvalue $\lambda_{\xi^-}$ vanishes at
$j=j_c$, upon which the saddle point (\ref{SP21}) ($\varepsilon=-$) merges
with the state (\ref{afm}) and thereby the partition functions
$\tilde{Z}_s$ and $Z_m$ tend to infinity. However, as $Z_m\propto
1/(j_c-j)$ increases much faster than $\tilde{Z}_s$, and
$|\kappa^-|\rightarrow 0$, $\Gamma_{j < j_c}^-$ tends to zero as
$j\rightarrow j_c$.
On the other hand, for $\varepsilon=+$, both $\lambda_{\xi^-}$ and
$\lambda_{\zeta^-}$ vanish leading to a divergent relaxation rate, since
now $\tilde{Z}_s$ diverges but $Z_m$ in Eq.~(\ref{Z_m})
remains finite for the metastable state (\ref{mfm}) is well defined.
Indeed, as $j\rightarrow j_c$, the relaxation rate
$\Gamma_{j < j_c}^+$ goes over to the result in Eq.~(\ref{Gamma_tsp_j>jc})
upon making the change $j\leftrightarrow j_c$, and taking account of the
symmetry factor.
In this case, the divergence at the point $j=j_c$ cannot be cut off
by expanding the energy beyond the $2^{nd}$ order and $\Gamma_{j <
j_c}^+$ is simply cut off at the point where it joins $\Gamma_{j > j_c}$
taking account of Eq.~(\ref{E6}).

In Fig.~\ref{rate} we plot ($ln$ of) the relaxation rate of the two-spin
system as defined in Eq. (\ref{Gamma_tsp}), under the condition of
IHD, in which Langer's approach is valid, that the reduced barrier height
$\beta\Delta e^{(0)}\gg 1$ and $\alpha\beta\Delta
e^{(0)}>1$ \cite{cofgarmcc01acp}. We also plot separately both relaxation rates
for $j\leq j_c$.
We see that the relaxation rate of the two-spin system contains two
unconnected branches corresponding to the two regimes, $j < j_c$ and
$j > j_c$, the bridging of which would require a more sophisticated
approach.
Fig.~\ref{rate} also shows that as $j$ increases, but $j \ll j_c$, the relaxation
rate $\Gamma_{j\leq j_c}^+$ decreases because the switching of the first
spin is hindered by the (ferromagnetic) exchange coupling. While $\Gamma_{j\leq
j_c}^-$ is an increasing function of $j$ with a faster rate, since now the
exchange coupling works in favor of the switching of the second spin.
This is also illustrated by the evolution of the energy barrier height in
Fig.~\ref{BH}.
As $j$ approaches $j_c$ from below, the relaxation rate $\Gamma_{j\leq
j_c}^+$ tends to $\Gamma_{j \geq j_c}$ because the respective saddle points
merge at $j=j_c$.
Whereas $\Gamma_{j\leq j_c}^-$ goes to zero since the corresponding saddle
point merges with the antiferromagnetic state that is no longer
accessible to the system.
For $j\geq j_c$, as $j$ increases the
minimum (\ref{sfm}), the metastable state (\ref{mfm}) and the saddle point
(\ref{SP0}) merge together, which means that the system is
found in an ``energy groove'' along the direction $\theta_1=\theta_2$
because the eigenvalue $\lambda_{\xi_-}$ corresponding to the mode
$\theta_1-\theta_2$ becomes very large, and thus the escape rate decreases
and eventually reaches the N\'eel-Brown value at large $j$.
\section{\label{sec:discussion}Discussion: What is wrong with the one-spin
approximation ?} %
We have seen that the magnetization of a nanoparticle can overcome the energy
barrier and thus reverse its direction, at least in two ways: either under
applied magnetic field which suppresses the barrier, or through thermally
activated statistical fluctuations.
Within the framework of the one-spin approximation, the
magnetization switching under applied field, at zero temperature, is well
described by the Stoner-Wohlfarth model.
At finite temperature and short-time scales, crossing of the energy barrier
activated by thermal fluctuations is described by the N\'eel-Brown model and its
extensions reviewed above.
Both of these models have been confirmed by experiments on individual
cobalt particles \cite{wer01acp}.
At finite temperature, but at quasi-equilibrium, the magnetization switching
occurs according to two distinct regimes. At very low temperature, this is
due to the coherent rotation of all spins, as in the Stoner-Wohlfarth model,
whereas at higher temperature, the magnetization switches by changing its
magnitude.
This results in a shrinking of the Stoner-Wohlfarth astroid as described by
the modified Landau theory \cite{kacgar01physa}, and confirmed by experiments
\cite{wer01acp}.
However, it is clear that the change of magnetization magnitude cannot be
explained in the framework of the one-spin approximation.
Indeed, it can only be understood as the result of a successive switching of
individual (or clusters of) spins inside the particle, which is necessarily a
multi-spin system.
Indeed, deviations from the single-spin approximation, and thereby
from both the Stoner-Wohlfarth and N\'eel-Brown models, have been observed in
metallic particles \cite{chesorklahad95prb}, \cite{resetal98prb}, and ferrite
particles \cite{kodber99prb}, \cite{ricetal91jap}.
These deviations have materialized in terms of the absence of magnetization
saturation at high fields, shifted hysteresis loops after cooling in field, and
enhancement at low temperature of the magnetization as a function of applied
field.
The latter effect has been clearly identified in dilute assemblies of maghemite
particles \cite{troetal00jmmm} of $4$ nm in diameter.
In addition, aging effects have been observed in cobalt single particles and
have been attributed to the oxidation of the sample surface into
antiferromagnetic CoO or NiO [see Ref.~\cite{wer01acp} and references
therein for a discussion of this issue].
It was argued that the magnetization reversal of a ferromagnetic particle with
antiferromagnetic shell is governed by two mechanisms that are supposed to be
due to the spin frustration at the core-shell interface of the particle.
On the other hand, according to M\"ossbauer spectroscopy some of the
above-mentioned novel features are most likely due to magnetic disorder at the
surface which induces a canting of spins, or in other words,
an inhomogeneous magnetic state inside the particle.

As argued earlier, understanding these effects requires the development of
microscopic theories capable of distinguishing and accounting for the various
crystallographic local environments that develop inside a nanoparticle and
on its surface.
Elaboration of such theories is faced with tough non-linear $N$-body problems
whose study can only be, in principle, efficiently performed with the help of
numerical approaches.
It is however desirable, and indeed even necessary, that the numerical
calculations be backed by analytical expressions, in some limiting cases at
least.

Accordingly, the study of section~\ref{sec:hk_tsp} has helped us understand the
effect of exchange coupling on the relaxation rate of the two-spin system.
The corresponding analytical results will be very
helpful in a generalization to multi-spin small particles at least for
small values of surface anisotropy and thereby small deviations from
collinearity, where it has been shown \cite{garkac03prl} that the surface
contribution to the macroscopic energy has a simple cubic anisotropy.
However, this generalization can only be performed using numerical
techniques. This is now attempted by the help of the ridge method
\cite{ioncar93jcp} for probing the potential energy surface and locating the
saddle points, and by the (Onsager-Machlup) path integrals \cite{ber98jmmm} for
determining the most probable paths connecting a metastable state to a more
stable state.
\section{acknowledgments}
I would like to thank my colleagues W.T. Coffey, D.A. Garanin, M.
Nogu\`es, E. Tronc, and W. Wernsdorfer for fruitful collaborations.

\end{document}